\documentclass[11pt, reqno,titlepage]{article}
\usepackage{graphicx,amsmath,amssymb,epsfig}
\usepackage{amsfonts}

\usepackage{hyperref}

\usepackage[
backend=biber,
style=apa,
sorting=nyt
]{biblatex}
\DeclareSourcemap{
  \maps[datatype=bibtex]{
    \map{
      \step[fieldset=issn, null]
      \step[fieldset=doi, null]
      \step[fieldset=url, null]
      \step[fieldset=urldate, null]
    }
  }
}

\newcommand{\citep}[2][]{\citeauthor{#2} (\citeyear[#1]{#2})}

\hypersetup{
    colorlinks=true,
    urlcolor=blue,
    linkcolor=blue,
    citecolor=blue
}

\newtheorem{theorem}{Theorem}

\newtheorem{corollary}{Corollary}

\newtheorem{definition}{Definition}

\newtheorem{lemma}{Lemma}

\newtheorem{proposition}{Proposition}

\newenvironment{varproof}[1][Proof]{\begin{trivlist}
\item[\hskip \labelsep {\bfseries #1}]}{\end{trivlist}}
\numberwithin{equation}{section}

\def\bkR{{\rm I\kern-.17em R}}
\def\uniset{{\rm 1\kern-.40em 1}}

\begin{document}

\title{\textbf{Matching with Trade-offs:}\\
Revealed Preferences over Competing Characteristics}
\author{Alfred Galichon\thanks{%
Economics Department, \'{E}cole polytechnique; e-mail:
alfred.galichon@polytechnique.edu} \and Bernard Salani\'e \thanks{%
Department of Economics, Columbia University; e-mail: bsalanie@columbia.edu.
} }
\date{First version dated December 6, 2008. The present version is of October 14, 2009
\thanks{%
The authors are grateful to Guillaume Carlier, Pierre-Andr\'e Chiappori,
Piet Gauthier, Jim Heckman, Guy Laroque, Rob Shimer as well as seminar
participants at Crest, Ecole Polytechnique, s\'eminaire Roy, University of
Chicago, and University of Alicante for useful comments and discussions. This paper is now superseded by `Cupids invisible hand' by the same authors.}.}
\maketitle

\begin{abstract}
We investigate in this paper the theory and econometrics of optimal
matchings with competing criteria. The surplus from a marriage match, for
instance, may depend both on the incomes and on the educations of the
partners, as well as on characteristics that the analyst does not observe.
Even if the surplus is complementary in incomes, and complementary in
educations, imperfect correlation between income and education at the
individual level implies that the social optimum must trade off matching on
incomes and matching on educations. Given a flexible specification of the
surplus function, we characterize under mild assumptions the properties of
the set of feasible matchings and of the socially optimal matching. Then we
show how data on the covariation of the types of the partners in observed
matches can be used to test that the observed matches are socially optimal
for this specification, and to estimate the parameters that define social
preferences over matches.

\noindent

{\footnotesize \ \textbf{Keywords}: matching, logit, generalized linear
models, revealed preferences, contingency tables. }

{\footnotesize \textbf{JEL codes}: C78, D61, C13. \vskip50pt }
\end{abstract}

\newpage



\section*{Introduction}

\begin{quote}
Louisa was naturally ill-tempered and cunning; but she had been taught to
disguise her real disposition, under the appearance of insinuating
sweetness, by a father who but too well knew that to be married would be the
only chance she would have of not being starved, and who flattered himself
that with such an extraordinary share of personal beauty, joined to a
gentleness of manners, and an engaging address, she might stand a good
chance of pleasing some young man who might afford to marry a girl without a
shilling.

Jane Austen, \emph{Lesley Castle} (1792).
\end{quote}

Starting with \citep{Becker:73}, most of the economic theory of
one-to-one matching has focused on the case when the surplus created by a
match is a function of just two numbers: the one-dimensional types of the
two partners. As is well-known, if the types of the partners are
one-dimensional and are complementary in producing surplus then the socially
optimal matches exhibit positive assortative matching. Moreover, the
resulting configuration is stable, it is in the core of the corresponding
matching game, and it can be implemented by the celebrated %
\citep{Gale-Shapley:62} deferred acceptance algorithm.

While this result is both simple and powerful, its implications are also
quite unrealistic. If we focus on marriage and type is education for
instance, then positive assortative matching has the most educated woman
marrying the most educated man, then the second most educated woman marrying
marrying the second most educated man, and so on. In practice the most
educated woman would weigh several criteria in deciding upon a match; even
in the frictionless world studied by theory, the social surplus her match
creates may be higher if she marries a man with less education but, say, a
similar income. Since income and education are only imperfectly correlated,
the optimal match must trade off assortative matching along these two
dimensions. This point is quite general: with multiple types, the stark
predictions of the one-dimensional case break down.

Empirical analysts of matching have long felt the need to accommodate the
imperfect assortative matching observed in the data, of course. This can be
done by introducing noise, in the form of heterogeneity in creation of
surplus that is unobserved by the analyst (see \citep{Choo-Siow:06}.)
Models with multidimensional types can also be estimated from the data, as
in \citep{CSTW:08}. But as far as we know, there has been little
theoretical work exploring the properties of optimal or equilibrium matches
in such models. We show in this paper that these properties can be summed up
in simple measures of covariation of types across partners; we analyze the
set of values of such measures that can be rationalized by a matching model;
and we show how to estimate this set from data and to test that the observed
matching is socially optimal\footnote{%
A word on terminology: like most of the literature, we call a
\textquotedblleft match\textquotedblright\ the pairing of two partners, and
a \textquotedblleft matching\textquotedblright\ the list of all realized
matches.}.

While we use the language of the economic theory of marriage in our
illustrations, nothing we do actually depends on it. The methods proposed in
this paper apply just as well to any one-to-one matching problem---or
bipartite matchings, to use the terminology of applied mathematics. In fact,
we can even extend them to problems in which the sets of partners are
determined endogenously---as with same-sex unions. This is investigated in
Section \ref{sect-extensions}, where we consider possible extensions of our
setting.

We do require, however, that utility be transferable across partners. Our
primitive function is indeed the surplus created by a match. We posit that
it is an unknown function of the types of the partners only, plus preference
shocks that are observed by all participants but not by the analyst---in the
nature of unobserved heterogeneity. When utility is transferable, all
optimal matchings must maximize the joint surplus; and so does the
equilibrium of the assignment game.

As is well-known, this model is too general to be empirically testable: even
without unobserved heterogeneity, any observed assignment can be
rationalized by a well-chosen surplus function. This is a consequence of a
more general theorem by \citep{Blair:84}. \citep{Echenique:08}
shows that on the other hand, some collections of matchings are not
rationalizable: if the analyst can observe identical populations on several
assignments, then these assignments must be consistent with each other in a
sense that his paper makes precise. But we are unlikely to have such data at
hand in general.

Relatedly, analysts sometimes observe several subpopulations which are
matching independently and yet have the same surplus function. %
\citep{Fox:identmatchinggames} shows that under a ``rank-order
condition'' on the unobserved heterogeneity, it is then possible to identify
several important features of the surplus function, and in particular how
important complementarity is on various dimensions.

While analyzing complementarity is also one of our goals here, many of the
applications we have in mind do not fit Fox's
assumption that there be enough variation across subpopulations with
identical surplus. Marriage markets, for instance, seem to be either so
disconnected that their surplus functions are unlikely to be similar, or too
connected to make it possible to ignore matching across markets. In this
paper, we will posit that we are only given data on one instance of a
matching problem, such as the marriage market in the US in the 1980s, or the
market for CEOs. Our data will consist of the values of the observable types
of both partners in each realized match, and of the types of unmatched
individuals. Since the optimal/equilibrium matching is determined on the
basis of both the observable and the (to us) unobservable types, we will
need to impose assumptions that allow us to integrate over the distribution
of the unobservable types in a manageable way. Our aim is to start from the
observable matching (the distribution of matches across observable types)
and to recover information on the observable surplus function (the average
surplus of matches for given observable types of both partners.)

To achieve this, we first resort to a separability assumption that rules out
interactions between the unobservable types of the partners in the surplus
form a match. This was used by~\citep{Choo-Siow:06}, and then
generalized by \citep{CSTW:08} who showed that the matching equilibrium
then boils down to a series of parallel discrete choice models. While this
is an important step on the way to a solution, the resulting model is still
too rich to be taken to the data. We need to restrict the distribution of
unobserved heterogeneity, and we do this by adopting again %
\citep{Choo-Siow:06}'s assumption that gives rise to multinomial
logits. Under these assumptions, we prove that the cross-differences of the
surplus function over observable types are nonparametrically identified from
the data. In particular, we can test for complementarities between any two
observable dimensions of the types of the partners, such as the education of
the wife and the income of the husband. We can also identify the relative
strengths of such complementarities across different dimensions.

If the analyst is lucky enough to have very rich data, then unobserved
heterogeneity is almost irrelevant and the observable matching maximizes the
observable surplus function. On the other hand, if data is so poor that
unobserved heterogeneity dominates, then the analyst should observe
something that, to him, looks like completely random matching. We show that
under our assumptions, this amounts to maximizing the mutual information of
the match distribution---a statistical object that here measures covariation
of partner types. Moreover, for any intermediate amount of unobserved
heterogeneity, the observable matching maximizes a straightforward linear
combination of the observable surplus and of mutual information.

This observation suggests a strategy: approximate the observable surplus
function with a linear expansion over some known basis functions, with
unknown \textquotedblleft assorting weights\textquotedblright . Then all
relevant information can be expressed in terms of the average values of
these basis functions across couples, and our results have a very neat
geometrical interpretation. Take the abstract space where each point
represents an hypothetical set of values for all the basis functions. All
feasible matchings generate points within a polytope in this space. We first
show that even with our restrictive assumptions, any point in this polytope
is rationalizable: if the variance of unobserved heterogeneity is
well-chosen, then there exist assorting weights for which the optimal
matching generates exactly these covariations. Fortunately, this combination
of heterogeneity and assorting weights is in general unique as we shall see:
this allows us to introduce several consistent and asymptotically normal
estimators of both the assorting weights and the variance of the unobserved
heterogeneity. Moreover, models without any unobserved heterogeneity can
only generate points on the boundary of the polytope, and so the homogeneous
model is testable.

This paper thus proves both a negative and a positive result. The negative
part is that even if we assume separable heterogeneity with a multinomial
logit structure, the model still cannot be rejected. The positive part is
that given any theory about the way the observable types enter the surplus
function (as embodied in a set of basis functions), we exhibit well-behaved
estimators of the unknown parameters; and we can quantify how much
heterogeneity is needed to rationalize the data. Moreover, our methods can
be used heuristically, to explore ways to understand what goes on in
matching markets---and how they change across time and space. Standard
statistical techniques could for instance be put to work to find the basis
functions that explain the largest share of the variation in the data. Such
a methodological stance is reminiscent of revealed preferences in consumer
theory; in fact the analogy is very sharp, as the underlying theoretical
structure is the same.

Our depiction of matching markets of course abstracts from many features of
real-world markets. We focus on static, frictionless markets, as in much of
the literature on marriage markets. Models of matching on job markets, for
instance, have on the whole adopted a much more dynamic perspective, in
which job flows in fact provide a lot of information on the underlying
parameters. In the applications we have in mind, the surplus function may
involve many more dimensions and we do not want to restrict it too much a
priori. The basic lack of identification mentioned at the beginning of this
introduction would become even more severe if we introduced dynamics or
frictions, unless these additional features are drastically simplified. We
leave this for further research. The paper also currently focuses on
discrete characteristics; we are exploring possible extensions to continuous
types.

Section~1 sets up the matching model we study in the paper, along with the
assumptions on the specification of the observable surplus and the process
that drives unobserved heterogeneity. In section~2 we build on these
assumptions to derive our main analytical results, and we give them a
geometric interpretation on section~3. Section~4 introduces our tests and
estimators and derives their asymptotic properties. We conclude by sketching
extensions of our methods.

Since much of what we do uses convexity, we recall some definitions and
basic results in Appendix A. All proofs are collected in Appendices B and C.
Finally, we should note that there are close parallels between the analysis
we develop in the present paper and familiar notions in thermodynamics and
statistical physics. E.g the social utility of a matching evokes (minus) the
internal energy of a physical system, and the standard error of unobservable
heterogeneity parallels its physical temperature. Since the analogy may
prove to be as useful to others as it was to us, we elaborate on it in
Appendix D.

\vspace{1cm}

\textbf{Summary of the notation used in the paper.} For the reader's
convenience, we regroup here the notation introduced in the text. We
consider matches between $N$ men and $N$ women. $\mathfrak{S}_{N}$ is the
set of permutations of $\left\{ 1,...,N\right\} $. A man has a full type $%
\tilde{x}=(x,\varepsilon )$, where the econometrician observes $x$ but not $%
\varepsilon $; we use $\tilde{y}=(y,\eta )$ for a woman. $x$ is a random
vector with distribution $P$, and $\tilde{x}$ is distributed according to $%
\tilde{P}$; we use $Q$ and $\tilde{Q}$ for a woman. We denote $\mathcal{M}%
\left( P,Q\right) $ the set of probability distributions with margins $P$
and $Q$; we use $\mathcal{M}(\tilde{P},\tilde{Q})$ for the full types. We
denote $P\otimes Q$ the product measure which matches men and women
randomly. A feasible matching generates a probability $\tilde{\Pi}\in
\mathcal{M}\left( \tilde{P},\tilde{Q}\right) $, which assesses the odds that
a man with full type $\tilde{x}$ is married to a woman with full type $%
\tilde{y}$. A man with full type $\tilde{x}$ and a woman with full type $%
\tilde{y}$ generate together a full surplus $\tilde{\Phi}\left( \tilde{x},%
\tilde{y}\right) $. We call $\Phi (x,y)=E\left( \tilde{\Phi}(\tilde{X},%
\tilde{Y})|X=x,Y=y\right) $ the observable surplus; in some of the paper we
take it to be the structural quadratic form $\Phi \left( x,y\right)
=x^{\prime }\Lambda y$.

\section{The Assignment Problem}

Throughout the paper, we assume that two subpopulations $M$ and $W$ of equal
size must be matched\; each man (as we will call the members of $M$) must be
matched with one and only one member of $W$ (we will call them women.) Thus
we do not model the determination of the unmatched population (the singles)
in this paper; we take it as data. We elaborate on this point in our
concluding remarks. Note also that we assumed bipartite matching: the two
subpopulations which define admissible partners are exogenously given. This
assumption can also be relaxed; see Section \ref{sect-extensions}.

Throughout the paper, we illustrate results on the education/income example
sketched in the Introduction, which we denote (ER).

\subsection{Population characteristics}

Each man $m$ has an $r$-dimensional type $x_{m}$ of observable
characteristics, and a vector of unobserved characteristics $\varepsilon_{m}
$. Denote $\tilde{x}_{m}=\left( x_{m},\varepsilon _{m}\right) $ the full
description of man $m$'s characteristics, which we call his full type. Each
woman $w$ similarly has an $s$-dimensional type $y_{w}$ of observed
characteristics, and a full type $\tilde{y}_{w}=\left( y_{w},\eta
_{w}\right) $.

We denote $\tilde{P}$ (resp.\ $\tilde{Q}$) the distribution of \textit{full}
types $\tilde{x}$ (resp.\ $\tilde{y}$) in the subpopulation $M$ (resp.\ $W$%
), and $P$ (resp.\ $Q$) the distribution of \textit{observable} types $x$
(resp.\ $y$.) Thus $P$ is a probability distribution on $\mathrm{I\kern%
-.17emR}^{r}$ and $Q$ is a distribution on $\mathrm{I\kern-.17emR}^{s}$. In
observed datasets we will have a finite number $N$ of men and women, so that
$P$ and $Q$ are the empirical distributions over the characteristics samples
of the men $\left\{ x_{1},...,x_{N}\right\} $ and the women $\left\{
y_{1},...,y_{N}\right\} $, respectively.

Take the education/income example: there $r=s=2$, the first dimension of
types is education $E\in \left\{ D,G\right\} $ (dropout or graduate), and
the second dimension is income class $R$, which takes values $1$ to $n_R$ . $%
P$ describes both the number of graduates among men and the distributions of
income among graduate men and among dropout men.

\subsection{Matching}

The intuitive definition of a matching is the specification of ``who marries
whom'': given a man of index $m\in \left\{ 1,...,N\right\} $, it is simply
the index of the woman he marries, $w=\sigma \left( m\right) \in \left\{
1,...,N\right\} $. Imposing that each man be married to one and only one
woman at a given time translates into the requirement that $\sigma $ be a
permutation of $\left\{ 1,...,N\right\} $, which we denote $\sigma \in
\mathfrak{S}_{N}$. This definition is too restrictive in so far as we would
like to allow for some randomization. This could arise because a given type
is indifferent between several partner types; or because the analyst only
observes a subset of relevant characteristics, and the unobserved
heterogeneity induces apparent randomness.

A \textit{feasible matching} (or \textit{assignment}) is therefore defined
in all generality as a joint distribution $\tilde{\Pi}$ over types of
partners $\tilde{X}$ and $\tilde{Y}$, such that the marginal distribution of
$\tilde{X}$ is $\tilde{P}$ and the marginal distribution of $\tilde{Y}$ is $%
\tilde{Q}$. We denote $\mathcal{M}\left( \tilde{P},\tilde{Q}\right) $ the
set of such joint distributions. Note that when $x$ and $y$ are univariate,
a feasible matching can be equivalently specified through a copula.

A matching is said to be \textit{pure} if all conditional distributions $%
\tilde{\Pi} \left( .|\tilde{x}\right) $ and $\tilde{\Pi} \left( .|\tilde{y}%
\right) $ are point mass distributions. In a pure matching $\tilde{\Pi} $,
there exists an invertible map $T\left( \tilde{x}\right) $ such that a man
with type $\tilde{x}$ almost surely marries a woman of type $\tilde{y}%
=T\left( \tilde{x}\right) $, and conversely, a woman with type $\tilde{y}$
almost surely marries a man of type $\tilde{x}=T^{-1}\left( \tilde{y}\right)
$. (Of course, in the discrete case this map can be represented as a
permutation on indices.)

In the education/income example (ER), a pure matching is described by $%
(2n_r)^2-1$ numbers; but the marginals impose $2(2n_r-1)$ constraints, so
that $(2n_r-1)^2$ numbers are to be determined.

\subsection{Surplus of a match}

The basic assumption of the model is that matching man $m$ of type $\tilde{x}%
_{m}$ and woman $w$ of type $\tilde{y}_{w}$\ generates a joint surplus $%
\tilde{\Phi}(\tilde{x}_{m},\tilde{y}_{w})$, where $\tilde{\Phi}$ is a
deterministic function. Along with most of the matching literature, we
assume that

\vspace{8mm}

\textbf{Assumption (O): Observability.} Each agent observes the full
characteristics $\tilde{x}$ and $\tilde{y}$ of all men and all women, but
the econometrician only observes the subvectors $x$ and $y$.

\vspace{8mm}

Assumption (O) rules out asymmetric information between participants in the
market, as the economics of matching with incomplete information is a
subject of its own. On the other hand, we do not need to assume full
information as the notation seems to imply: $\tilde{\Phi}$ could for
instance be reinterpreted as the expectation of a random variable
conditional on $\tilde{x},\tilde{y}$, as long as all participants evaluate
it in the same way.

Given Assumption (O), we need to define the \textit{observable surplus} as
the best predictor of $\tilde{\Phi}(\tilde{x},\tilde{y})$ conditional on $x$
and $y$, that is
\begin{equation*}
\Phi (x,y)=E\left[ \tilde{\Phi}(\tilde{X},\tilde{Y})|X=x,Y=y\right]
\end{equation*}%
and we can write the decomposition%
\begin{equation*}
\tilde{\Phi}(\tilde{x},\tilde{y})=\Phi (x,y)+k \left( \tilde{x},\tilde{y}%
\right)
\end{equation*}%
where $k \left( \tilde{x},\tilde{y}\right) $ is the \textit{idiosyncratic
surplus}.

Following the insight of \citep{Choo-Siow:06}, formalized by %
\citep{CSTW:08}, we now assume:

\textbf{Assumption (S): Separability}. Let $\tilde{x}$ and $\tilde{x}%
^{\prime}$ have the same observable type: $x=x^{\prime}$. Similarly, let $%
\tilde{y}$ and $\tilde{y}^{\prime}$ be such that $y=y^{\prime}$. Then
\begin{equation*}
\tilde{\Phi}(\tilde{x},\tilde{y})+\tilde{\Phi}(\tilde{x}^{\prime},\tilde{y}%
^{\prime})= \tilde{\Phi}(\tilde{x},\tilde{y}^{\prime})+\tilde{\Phi}(\tilde{x}%
^{\prime},\tilde{y}).
\end{equation*}
While much of the literature on matching emphasizes complementarity,
assumption (S) in fact requires that conditional on observable types, the
surplus exhibit no complementarity across unobservable types.

It is easy to see that imposing assumption (S) is equivalent to requiring
that the idiosyncratic surplus from a match must be additively separable, in
the following sense:
\begin{equation*}
k\left( \tilde{x},\tilde{y}\right) =\chi \left( \tilde{x},y\right) +\xi
\left( \tilde{y},x\right) ,
\end{equation*}%
where $\chi $ and $\xi $ are two deterministic functions and
\begin{equation*}
E(\chi (\tilde{X},Y)|X=x,Y=y)=E(\xi (\tilde{Y},X)|X=x,Y=y)=0.
\end{equation*}

Then the surplus function $\tilde{\Phi}$ can be rewritten as%
\begin{equation*}
\tilde{\Phi}(\tilde{x},\tilde{y})=\Phi (x,y)+\chi \left( \tilde{x},y\right)
+\xi \left( \tilde{y},x \right) .
\end{equation*}
Note that the model is invariant if one rescales the three terms on the
right-hand side by the same positive constant. Later on we will normalize
these three components.

As proved in \citep{CSTW:08}, assumption (S) implies that at the
optimum (or equilibrium), a given individual (say, a man $\tilde{x}$) has a
preference $\xi \left( \tilde{x},y \right)$ for a particular class of
observable characteristics (say $y$), but he is indifferent between all
partners which have the same $y$ but a different $\eta$.

In fact, the optimal matching is characterized by two functions of
observable characteristics $U(x,y)$ and $V(x,y)$ that sum up to $\Phi(x,y)$
such that if a man $\tilde{x}=(x,\varepsilon)$ is matched with a woman of
characteristics $\tilde{y}=(y,\eta)$, he will get utility
\begin{equation*}
U(x,y)+\chi(\tilde{x},y)
\end{equation*}
while his match gets utility
\begin{equation*}
V(x,y)+\xi(\tilde{y},x).
\end{equation*}
\citep{CSTW:08} showed that given assumption (S), the matching problem
boils down to a set of discrete choice models for each type of man and of
woman: for instance, man $\tilde{x}$ is matched in equilibrium to a woman $%
\tilde{y}$ whose observable type $y$ maximizes
\begin{equation*}
U(x,y)+\xi(\tilde{x},y)
\end{equation*}
over all values in the support of $Q$.

While this is already quite useful, we need to add more restrictions on the
specification of the components of the idiosyncratic surplus $\chi \left(
\tilde{x},y \right) $ and $\xi \left( \tilde{y},x\right) $.

\subsection{Specifying the idiosyncratic surplus}

Following \citep{Choo-Siow:06} and \citep{CSTW:08}, we introduce
the following assumption\footnote{%
We define the scale factor to be 1 for the standard Gumbel, which has
variance $\pi ^{2}/6$; thus e.g. $\chi $ has variance $\sigma _{1}^{2}\pi
^{2}/6$.}:

\textbf{Assumption GUI: Gumbel Unobserved Interactions.} It is assumed that:

- There are an infinite number of individuals with a given observable type
in the population

- Fix the observable characteristics $x$ of a man, and let $\left(
y_{1}^{\ast },...,y_{T_y}^{\ast }\right) $ be the possible values of the
observable characteristics of women. Then the vector of preference shocks $%
\chi \left(x,\varepsilon,y_{1}^{\ast }\right) ,...,\chi
\left(x,\varepsilon,y_{T_{y}}^{\ast }\right) $ are distributed as $T_{y}$
independent and centered Gumbel random variables with scale factor $\sigma
_{1}$;

similarly,

- Fix the observable characteristics $y$ of a man, and let $\left(
x_{1}^{\ast },...,x_{T_x}^{\ast }\right) $ be the possible values of the
observable characteristics of men. Then the vector of preference shocks $\xi
\left(y,\eta,x_{1}^{\ast }\right) ,...,\xi \left(y,\eta,x_{T_{x}}^{\ast
}\right) $ are distributed as $T_{x}$ independent and centered Gumbel random
variables with scale factor $\sigma _{2}$.

$\blacksquare $

In short: men of a given observable type have conditionally Gumbel iid draws
of the $\chi $'s for different individuals; and conversely for women of a
given observable type.

We use (GUI) for the Independence of Irrelevant Alternatives property:
without it, the odds ratio of the probability that a man with observable
type $x$ ends up in a match with a woman of observable type $y$ rather than
with $z$ would also depend on the types of other women, and the model would
become unmanageable.

(GUI) underlies the standard multinomial logit model of discrete choice. It
has well-known limitations, one of which is that it does not extend directly
to continuous choice. We are currently exploring alternative specifications
that would allow us to deal with continuous characteristics; but at this
stage, we assume

\textbf{Assumption (DD):} The distributions of observed types $P$ and $Q\ $%
are discrete, with probability mass functions $p\left( x\right) $ and $%
q\left( y\right) $. $\blacksquare $

In the (ER) example for instance, $p(D,3)$ is the proportion of men who are
dropouts and whose income lies in class 3. For simplicity, we now denote $%
i_P=1,\ldots,n_P$ the possible values of types of men, and $i_Q=
1,\ldots,n_Q $ for women.

\subsection{Specifying the observable surplus}

We now introduce sets of assumptions on the observable surplus ranging from
non-restrictive (NPOI\ below, suited for nonparametric identification) to
more restrictive (SLOI\ below, convenient for a more concise analysis).

Let us first impose a normalization convention on the observable surplus.
Notice that the optimal matching (but not the value of the social surplus)
is left unchanged if we add an additively decomposable function $f\left(
x\right) +g\left( y\right) $ to $\Phi \left( x,y\right) $. Therefore,
without any loss of generality, we impose some identifying restriction on $%
\Phi $, using the two-way ANOVA decomposition, accoding to which any vector $%
\Phi \left( x,y\right) $ admits the following orthogonal decomposition in $%
L^{2}\left( \pi \right) $ as%
\begin{equation*}
\Phi \left( x,y\right) =\bar{\Phi}\left( x,y\right) +f\left( x\right)
+g\left( y\right) +c
\end{equation*}%
where $E_{p}\left[ f\left( X\right) \right] =E_{q}\left[ g\left( Y\right) %
\right] =0$ and $E\left[ \bar{\Phi}\left( X,Y\right) |X\right] =E\left[ \bar{%
\Phi}\left( X,Y\right) |Y\right] =0$. We shall therefore often take the
following convention when using a nonparametric approach:

\textbf{Convention (ZMOI): Zero-mean Observable interactions.} The
observable surplus satisfies%
\begin{equation*}
E\left[ \Phi \left( X,Y\right) |X\right] =E\left[ \Phi \left( X,Y\right) |Y%
\right] =0.
\end{equation*}

\bigskip

It will sometimes be useful to assume more structure on the function $\Phi $
(the observable joint surplus.) To do this, we consider $K$ given \emph{%
basis assorting functions} $\phi ^{1}(x,y),...,\phi ^{K}(x,y)$ whose values
are interpreted as the utility benefit of interaction between type $x$ and
type $y$. Given \emph{assorting weights} $\Lambda \in \mathbb{R}^{K}$, we
focus on \emph{\ observable surplus functions} $\Phi _{\Lambda }\left(
x,y\right) $ which are linear combinations of the basis assorting functions
with weights $\Lambda $. That is,

\textbf{Assumption (SLOI): Semilinear Observable Interactions.} The
observable surplus function can be written%
\begin{equation}
\Phi _{\Lambda }(x,y)=\sum_{k=1}^{K}\Lambda _{k}\phi ^{k}(x,y)
\label{generalV}
\end{equation}%
where the sign of each $\Lambda _{k}$ is unrestricted.$\blacksquare $

\bigskip

Note that in the discrete case which we restrict to in this paper, this
general form is absolutely \emph{not} restrictive. Indeed, one can choose $%
K=T_{x}\times T_{y}$ and chose $\phi ^{ij}\left( x,y\right) =1_{\left\{
x=x_{i},y=y_{j}\right\} }$ so that $\phi ^{ij}\left( x,y\right) $ captures
interaction between observable man type $x_{i}$ and observable woman type $%
y_{j}$. We shall refer to this specification as the:

\textbf{Specification (NPOI): Nonparametric Observable Interactions.} The
observable surplus function is expanded in all generality
\begin{equation}
\Phi _{\Lambda }(x,y)=\sum_{i=1}^{T_{x}}\sum_{j=1}^{T_{y}}\Lambda
_{ij}1_{\left\{ x=x_{i},y=y_{j}\right\} }.  \label{fullyNonparam}
\end{equation}%
in which case social weight $\Lambda _{ij}$ coincide with $\Phi \left(
x_{i},y_{j}\right) $.$\blacksquare $

However we favor parsimonious models for the sake of analysis, so in general
we shall only assume (SLOI), unless explicitely stated.

\bigskip

To return to the education/income example (ER): we could for instance assume
that a match between man $m$ and woman $w$ creates a surplus that depends on
the similarity of the partners in both education and income dimensions. The
corresponding specification would be (with education levels $E=(D,G)$ coded
as $(0,1)$):
\begin{equation*}
\Phi(x_{m},y_{w})=\sum_{e_m=0,1;e_w=0,1}\Lambda _{e_m,e_w}\mathrm{1\kern%
-.40em 1}(E_{m}=e_m,E_{w}=e_w) \;
+\sum_{i=1,\ldots,n_r;j=1,\ldots,n_r}\Lambda _{ij}\mathrm{1\kern-.40em 1}%
(R_m=i,R_w=j).
\end{equation*}
This specification only has $(n^2_r+4)$ parameters, while an unrestricted
specification would have $4n_r^2$. Such an unrestricted specification would
for instance allow the effect of matching partners in income class 3 to
depend on both of their education levels.

An even more restrictive, ``diagonal'' specification would be
\begin{equation*}
\Phi(x_{m},y_{w})=\sum_{e=0,1}\Lambda^E_e\mathrm{1\kern-.40em 1}%
(E_{m}=E_w=e) \; +\sum_{i=1,\ldots,n_r}\Lambda^R _{i}\mathrm{1\kern-.40em 1}%
(R_m=R_w=i).
\end{equation*}

In this last form, it is clear that the relative importance of the $\Lambda $%
's reflects the relative importance of the criteria. Thus $\Lambda _{i}^{R}$
measures the preference for matching partners who are both in income class $%
i $, while $\Lambda _{0}^{E}$ measures the preference for matching dropouts.
The relative values of these numbers indicate how social preferences value
complementarity of incomes of partners more, relative to complementarity in
educations. We will not need to assume such a diagonal structure in the
following, although our results easily specialize to this case.

\subsection{Summary: the model specification}

Under assumptions (O), (S), (SLOI) and (GUI), the model is fully
parametrized; its parameters can be collected in a vector
\begin{equation*}
\theta =(\Lambda ,\sigma _{1},\sigma _{2}),
\end{equation*}%
where $\Lambda $ is the assorting weight matrix, and $\sigma _{1}$ (resp.\ $%
\sigma _{2}$) is the scale factor of the unobservable characteristics of the
men (resp.\ of women). Without loss of generality, all components of $\theta
$ can be multiplied by any positive number; hence we shall need to impose
some normalization on $\theta$.

Most of the results in the next section in fact only require assumptions
(O), (S) and (GUI), with a general function $\Phi(x,y)$. In this case $%
\theta $ is just $(\Phi,\sigma_1,\sigma_2)$, and again it is defined up to a
scale factor.

As we will see, the \emph{total heterogeneity\/} $(\sigma_1+\sigma_2)$ plays
a key role in our results; thus we introduce a specific notation for it:
\begin{equation*}
\sigma=\sigma_1+\sigma_2.
\end{equation*}

\section{Solving for the Optimal Matching\label{sec:optimal}}

In this section we only assume (O), (S), and (GUI), and we consider the
problem of optimal matching:
\begin{equation}
\mathcal{W}(\theta )=\sup_{\tilde{\Pi}\in \mathcal{M}\left( \tilde{P},\tilde{%
Q}\right) }E_{\tilde{\Pi}}\left[ \tilde{\Phi}\left( \tilde{X},\tilde{Y}%
\right) \right] .  \label{optim}
\end{equation}

Our modeling strategy in this section and the next is to assume that the
number of men and women in the population is large enough that averages can
be replaced with expectations. When we describe our estimators in section~%
\ref{sec:inference}, we of course take into account the fact that we only
have a finite sample.

\subsection{The heterogeneous model}

Let us provide some intuition before we state a formal theorem. Under (O),
(S) and (GUI), standard formul\ae\ of the multinomial logit model give the
expected utility of a man of observable type $x$ at the optimal matching:
\begin{equation*}
E\left[ \max_{y}\left( U(x,y)+\chi (\tilde{X},y)\right) |X=x\right] =\sigma
_{1}\log \sum_y \exp \left( U(x,y)/\sigma _{1}\right).
\end{equation*}%
Therefore the expected social surplus from the optimal matching is simply%
\footnote{%
Since this formula may not be entirely transparent, we develop one term
below:
\begin{equation*}
E_{P}\log \sum_y \exp (U(X,y)/\sigma _{1})=\sum_x p(x)\log \sum_y \exp
\left( U(x,y)/\sigma _{1}\right) .
\end{equation*}%
} (adding the equivalent formula for women of observable type $y$):
\begin{equation*}
\sigma _{1}E_{P}\log \sum_y\exp (U(X,y)/\sigma _{1})+\sigma _{2}E_Q\log
\sum_x\exp (V(x,Y)/\sigma _{2}).
\end{equation*}

Now recall that $U(x,y)$ is the mean utility of a man with observable type $%
x $ who ends up being matched to a woman with observable type $y$ at the
optimum. As in the general development of the theory of matching, $U$ is the
value of the multiplier of the population constraints; and as such, it
(along with $V$) is the unknown function in the dual program in which the
expression for the social surplus above is minimized over all $U,V$ such
that $U+V\geq \Phi$. We now state this as a theorem (proved in the Appendix):

\begin{theorem}[Social welfare-primal version]
\label{theorem-socialw-primalp} Assume (O), (S), (GUI) and (DD). Then
\begin{equation}
\mathcal{W}(\theta )=\inf_{(U,V)\in A}\left( \sigma _{1}E_{P}\log \sum_y
\exp (U(X,y)/\sigma _{1})+\sigma _{2}E_{Q}\log \sum_x \exp (V(x,Y)/\sigma
_{2})\right)  \label{PrimalPT}
\end{equation}%
where the constraint set $A$ is defined by the inequalities
\begin{equation*}
\forall x,y,\;U(x,y)+V(x,y)\geq \Phi \left( x,y\right) .
\end{equation*}
\end{theorem}

At an optimal matching, men with observable type $x$ will be found in
matches with women with observable types $y$ such that $U(x,y)+V(x,y)=\Phi
\left( x,y\right) $. The expected utility of men with observable type $x$
matched with women of observable type $y$ is $U(x,y)$.

This theorem also has a primal version, of course. While deriving it takes a
bit more work (again, see the Appendix), the intuition is simple. First, if
there were no unobserved heterogeneity (with $\sigma$ close to zero) the
optimal matching would coincide with the optimal observable matching $\Pi $,
which solves
\begin{equation*}
\mathcal{W}(\theta )=\sup_{\Pi \in \mathcal{M}\left( P,Q\right) }E_{\Pi
}\Phi \left( X,Y\right) .
\end{equation*}%
Going to the polar opposite, in the limit when $\sigma$ goes to infinity
only unobserved heterogeneity would count; and since it is just noise, the
optimal matching would simply assign partners randomly, yielding the product
measure $P\otimes Q$.

As it turns out, when $\sigma $ takes any intermediate value the optimal
matching maximizes a weighted sum of these two extreme cases:

\begin{theorem}[Social welfare-dual version]
Under the assumptions of Theorem~\ref{theorem-socialw-primalp} \label%
{theorem-socialw-primal}
\begin{equation}
\mathcal{W}(\theta )=\sup_{\Pi \in \mathcal{M}\left( P,Q\right) }\left(
\sum_{x,y}\pi (x,y)\Phi \left( x,y\right) -\sigma I\left( \Pi \right)
\right) +\sigma _{1}S(Q)+\sigma _{2}S(P),  \label{DualPT}
\end{equation}%
where $S\left( P\right) $ and $S\left( Q\right) $ are the \emph{entropies}
of $P$ and $Q$ given by%
\begin{equation*}
S(P)=-\sum_{x}p(x)\log p(x);\text{ }\mbox{ and }S(Q)=-\sum_{y}p(y)\log p(y);
\end{equation*}%
and $I(\Pi )$ is the \emph{mutual information\/}of joint distribution $\Pi $
, given by%
\begin{equation*}
I(\Pi )=\sum_{x,y}\pi (x,y)\log \frac{\pi (x,y)}{p(x)q(y)}.
\end{equation*}
\end{theorem}

The mutual information $I\left( \Pi \right) $ is nothing else than the
Kullback-Leibler divergence of $\Pi $ from the independent product $P\otimes
Q$ to $\Pi $. Recall two important information-theoretic properties of $I$:

\begin{enumerate}
\item The map $\pi \rightarrow I\left( \pi \right) $ is strictly convex.

\item One has
\begin{equation*}
\forall \Pi \in \mathcal{M}\left( P,Q\right) ,~S\left( P\right) +S\left(
Q\right) \geq I\left( \Pi \right) \geq 0
\end{equation*}%
the left handside becoming an equality in particular in the case of a pure
matching, and the right handside inequality becoming an equality in the case
where $\Pi =P\otimes Q$, as we shall see below.
\end{enumerate}

Mutual information is a measure of the covariation of types $x$ and $y$. Now
$P\otimes Q$ is the independent product of $P$ and $Q$, which corresponds to
a completely random matching $\Pi =P\otimes Q$. Thus a large positive $%
I\left( \Pi \right) $ indicates that the matching $\Pi $ induces strong
correlation across types; $I(\Pi )=S\left( P\right) +S\left( Q\right) $ if
and only if $\Pi =P\otimes Q$. If $\sigma $ is very large then the Theorem
suggests that $I(\Pi )$ should be minimized, which can only occur for the
independent matching $\Pi =P\otimes Q$; whereas if $\sigma $ is negligible
then $\Pi $ should be chosen so as to maximize the expected \emph{%
observable\/} surplus $E_{\Pi }\Phi (X,Y)$. This corroborates the intuition
given earlier.

\bigskip

Now the optimal matchings coincide with the solutions to this maximization
problem. Since we only observe the realized $\Pi $ over observable
variables, Theorem~\ref{theorem-socialw-primal} defines the empirical
content of the model: a combination of the parameters $\theta =(\Phi ,\sigma
_{1},\sigma _{2})$ is identified if and only if the solution $\Pi $ depends
non-trivially on it.

We already knew that $\theta$ can be rescaled by any positive constant
without altering the solution. We can now go one step further: while all
components of $\theta $ figure in this theorem, $\sigma _{1}$ and $\sigma
_{2}$ only enter through their sum $\sigma$. Thus and as announced, $\sigma
_{1}$ and $\sigma _{2}$ are not separately identified.

Accordingly, we redefine the parameter vector $\theta $ as%
\begin{equation*}
\theta =\left( \Phi ,\sigma \right),
\end{equation*}
or $\theta=(\Lambda,\sigma)$ under (SLOI).

\subsection{The homogeneous limit}

In this section we consider the limit behavior of our model when $\sigma$
goes to zero, so that unobservable heterogeneity vanishes. We denote
\begin{equation*}
\mathcal{W}_0(\Phi) \equiv \mathcal{W}(\Phi,0).
\end{equation*}

By taking the limit in Theorem \ref{theorem-socialw-primalp}, we obtain:

\begin{theorem}[Homogeneous social welfare]
\label{thm-socialw-homo} Assume (O) and (DD); then

a) The value of the social optimum when $\theta =\left( \Phi ,0\right) $ is
given both by
\begin{equation}
\mathcal{W}_{0}\left( \Phi \right) =\max_{\Pi \in \mathcal{M}\left(
P,Q\right) }\sum_{x,y}\pi (x,y)\Phi \left( x,y\right) ,  \label{Primal0}
\end{equation}%
and by%
\begin{equation}
\mathcal{W}_{0}\left( \Phi \right) =\inf_{(u,v)\in A^{0}}\left(
\sum_{x}p(x)u\left( x\right) +\sum_{y}q(y)v\left( y\right) \right)
\label{Dual0}
\end{equation}%
where the constraint set $A^{0}$ is given by
\begin{equation*}
\forall x,y,\;\;u(x)+v(y)\geq \Phi \left( x,y\right) ;
\end{equation*}%
A matching $\left( X,Y\right) \sim \Pi $ is optimal for $\Phi $ if and only
if the equality
\begin{equation*}
u\left( X\right) +v\left( Y\right) =\Phi \left( X,Y\right)
\end{equation*}%
holds $\Pi $-almost surely, where $u$ and $v$ solve the optimization problem
(\ref{Dual0}).
\end{theorem}

Thus in the limit we recover the standard primal and dual formulation of the
matching problem; since all men with observable characteristics $x$ have the
same tastes, they all obtain the same utility at the optimum and $U(x,y)$
becomes a function of $x$ only, which we denoted $u(x)$ above; and this is
just the Lagrange multiplier on the population constraint
\begin{equation*}
\sum_{y}\pi (x,y)=p(x)
\end{equation*}%
which is implicit in the notation $\Pi \in \mathcal{M}(P,Q)$.

\section{Qualitative properties of the optimum}

In this section we first introduce the various statistics on which our
analysis shall rest. We then provide comparative statics which help
understanding the influences on the model parameters on these statistics;
last, we study the influence on qualitative properties of the equilibria
such as uniqueness and purity of the equilibria.

\subsection{Matching summaries}

\textbf{Feasible summaries.} Recall that under (SLOI), there exists an
unknown vector $\Lambda $ such that the observable surplus function takes
the form
\begin{equation*}
\Phi (x,y)=\sum_{k=1}^{K}\Lambda _{k}\phi ^{k}(x,y)
\end{equation*}%
with known basis functions $\phi ^{k}$. Now consider a hypothetical matching
$\Pi $; under this matching, the basis functions have expected values
\begin{equation*}
C^{k}(\Pi )=\sum_{x,y}\pi (x,y)\phi ^{k}\left( x,y\right) .
\end{equation*}%
We call each $C^{k}$ a \emph{covariation}. Take the (ER) example; then

\begin{itemize}
\item $C^1$ is the proportion of matches among graduate partners under $\Pi$

\item $C^2$ is the expected income of a graduate man's wife multiplied by
the proportion of graduate men; $C^3$ is defined similarly

\item and $C^4$ is the expected product of the partners' incomes.
\end{itemize}

Random matching, as represented by $\Pi _{\infty }=P\otimes Q$, plays a
special role in our analysis, as it obtains in the limit when heterogeneity
becomes very large. We denote the corresponding covariations as $C_{\infty
}^{k}$. At the polar opposite is the matching $\Pi _{0}$ that obtains in the
homogenous limit $\sigma =0$; we denote the implied covariations $%
C_{0}^{k}(\Lambda )$. Note that $C_{\infty }$ does not depend on $\Lambda $,
but $C_{0}$ does.

We know from Theorem~\ref{theorem-socialw-primal} that under (SLOI), the
observable optimal matching $\Pi $ maximizes
\begin{equation*}
\Lambda \cdot C(\Pi )-\sigma I(\Pi ).
\end{equation*}%
Thus the vector $(C(\Pi ),I(\Pi ))$ summarizes all the relevant information
about matching $\Pi $. We call each such vector a \emph{matching summary};
matching summary vectors are set in \emph{summary space}, which is a subset%
\footnote{%
Remember that $I(\Pi )\geq 0$ for any feasible matching.} of $\mathrm{I\kern%
-.17emR}^{K+1}$.

Given an observed matching, it is of course very easy to estimate the
associated covariation vector and mutual information. Again, the model is
scale-invariant and we may impose an arbitrary normalization on $\theta
=(\Lambda ,\sigma )$. For that purpose we choose a vector $C^{\ast }$ and we
impose $\Lambda \cdot C^{\ast }=1$. Later we make the choice of $C^{\ast }$
more specific.

\bigskip

Given population distributions $P$ and $Q$, we define the \emph{set of
feasible summaries\/}$\mathcal{F}$ as the set of summary vectors $(C,I)$
that are generated by some feasible matching $\pi \in \mathcal{M}(P,Q)$,
that is%
\begin{equation*}
\mathcal{F=}\left\{ \left( C,I\right) \in \mathbb{R}^{K}\times \left[
0,S\left( P\right) +S\left( Q\right) \right] :\exists \Pi \in \mathcal{M}%
(P,Q),~C^{k}=C^{k}\left( \Pi \right) ,~I=I\left( \Pi \right) \right\}
\end{equation*}

Similarly, define the \emph{covariogram} $\mathcal{F}_{c}$ as the set of
covariations $C$ that are implied by some feasible matching; that is,%
\begin{equation*}
\mathcal{F}_{c}\mathcal{=}\left\{ C:\exists \Pi \in \mathcal{M}%
(P,Q),~C^{k}=C^{k}\left( \Pi \right) \right\} .
\end{equation*}

\bigskip

\begin{figure}[tbp]
\begin{center}
\includegraphics[width=\textwidth]{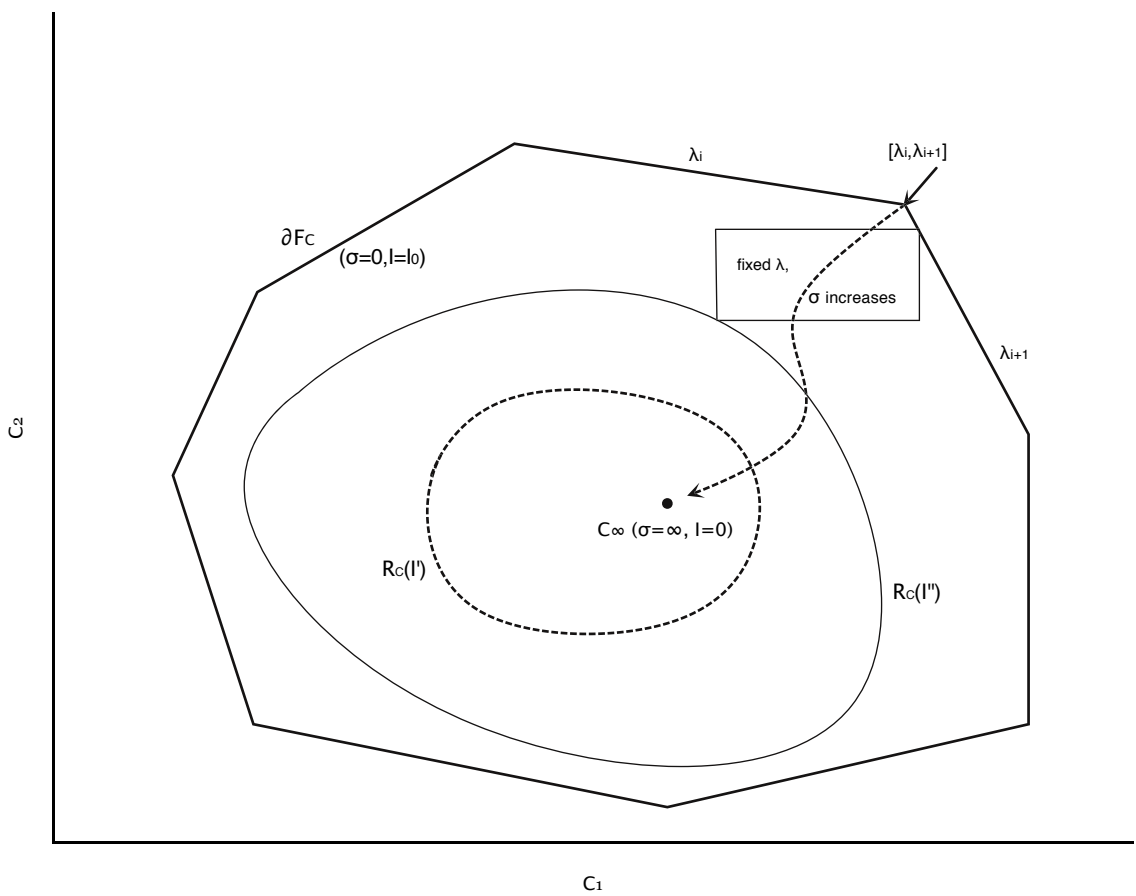}
\end{center}
\caption{The covariogram and related objects}
\label{Figure:CovarD}
\end{figure}

Covariograms provide us with a nice graphical representation of the
properties of a matching. Figure~\ref{Figure:CovarD} illustrates their
relevant properties, and the reader should refer to it as we go along. To
fit it within two dimensions, we assume that there are only two basis
functions; e.g. in the (ER) example we could have
\begin{equation*}
\Phi (E_{m},E_{w},R_{m},R_{w})=\Lambda _{1}\mathrm{1\kern-.40em1}%
(E_{m}=E_{w})+\Lambda _{2}\mathrm{1\kern-.40em1}(R_{m}=R_{w}),
\end{equation*}%
so that $\Lambda _{1}$ (resp.\ $\Lambda _{2}$) measures the preference for
assortative matching on educations (resp.\ income classes.)

\begin{proposition}
\label{prop-convexFeasible}Under (O), (S), (GUI) and (SLOI), the sets $%
\mathcal{F}$ and $\mathcal{F}_{c}$ are nonempty closed convex sets, and
their support functions are $\mathcal{W}\left( \Lambda ,\sigma \right) $ and
$\mathcal{W}\left( \Lambda ,0\right) $, respectively.
\end{proposition}

As will soon become clear, the boundaries of the convex sets $\mathcal{F}$
and $\mathcal{F}_{c}$ have special significance in our analysis. For now,
let us simply note that the boundary of $\mathcal{F}_{c}$ exhibits kinks
when these distributions of characteristics are discrete---which is always
the case in our setting. The reason for these kinks is that in the discrete
case, the optimal matching for homogenous types is generically stable under
a small perturbation of the assorting weights $\Lambda $; starting from
almost every\ $\Lambda $'s, a small change in $\Lambda $ leaves covariations
unchanged. Any such value of $\Lambda $ generates a covariation vector on a
vertex of the polytope. On the other hand, there exist a finite number of
values of $\Lambda $ where the optimal matching problem has multiple
solutions, with corresponding multiple covariations; each such value of $%
\Lambda $ generates a facet of the polytope. This is shown on Figure~\ref%
{Figure:CovarD} with all $\lambda =\Lambda _{2}/\Lambda _{1}$ in an interval
$[\lambda _{i},\lambda _{i+1}]$ generating the same covariations in the
homogeneous case. Remarkably, such kinks disappear as soon as there is
enough positive amount of heterogeneity; we will come back to this in
section~\ref{smoothsection}.

\subsection{Rationalizable boundary}

The previous discussion suggests an intimate connection between the
boundaries of the sets described above and optimal matchings. To make this
clear, we now define the \textit{set of rationalizable summaries} $\mathcal{R%
}$ as the set of $(K+1)$-uples $\left( C,I\right) $ such that $C$ and $I$
are covariations and mutual information corresponding to an optimal matching
$\Pi \in \mathcal{M}(P,Q)$ for some parameter values $\left( \Lambda ,\sigma
\right) $:%
\begin{equation*}
\mathcal{R=}\left\{ \left( C,I\right) \in \mathcal{F}:\exists \left( \Lambda
,\sigma \right) \in \mathbb{R}^{K}\times \left[ 0,S\left( P\right) +S\left(
Q\right) \right] ,~\Lambda \cdot C-\sigma I=\mathcal{W}\left( \Lambda
,\sigma \right) \right\} .
\end{equation*}%
Obviously, rationalizable summaries are feasible and $\mathcal{R}\subset
\mathcal{F}$. This definitions allow us to state that rationalizable
summaries and extreme feasible summaries coincide. Or, to put it more
formally:

\begin{proposition}
\label{prop-frontierFeasible}$\mathcal{R}$ is the frontier of $\mathcal{F}$\
in $\mathbb{R}^{K}\times \left[ 0,S\left( P\right) +S\left( Q\right) \right]
$.
\end{proposition}

\bigskip

\textbf{Mutual information level sets.} Now consider a covariation vector $C$
in the covariogram $\mathcal{F}_{c}$, and define the \emph{rationalizing
mutual information}
\begin{equation*}
I_{r}\left( C\right) :=\sup \left\{ I\in \left[ 0,S\left( P\right) +S\left(
Q\right) \right] :\left( C,I\right) \in \mathcal{R}\right\} ;
\end{equation*}%
clearly from the definition of $\mathcal{R}$ and positive homogeneity of $%
\mathcal{W}$, we see that the \emph{implicit mutual information function}
\begin{equation}
I_{r}\left( C\right) =\sup_{\lambda }\left\{ \lambda \cdot C-\mathcal{W}%
\left( \lambda ,1\right) \right\}  \label{IrC}
\end{equation}%
so $I_{r}\left( C\right) $ is the Legendre-Fenchel transform of $\mathcal{W}%
\left( \Lambda ,1\right) $ which is strictly convex; in particular $%
I_{r}\left( C\right) $ is a $C^{1}$ function. Conversely, for any mutual
information $I\geq 0$ we define and the \emph{set of rationalizable
covariations\/} by
\begin{equation*}
\mathcal{R}_{c}\left( I\right) =\left\{ C\in \mathbb{R}^{K}:\exists I\in %
\left[ 0,S\left( P\right) +S\left( Q\right) \right] ,~\left( C,I\right) \in
\mathcal{R}\right\} =I_{r}^{-1}\left( \left\{ I\right\} \right) .
\end{equation*}

It follows directly from the convexity of $I_{r}$ that $\mathcal{R}%
_{c}\left( I\right) $ is the boundary of the set $I_{r}^{-1}\left( \left[ 0,I%
\right] \right) $, which is convex and increasing (for inclusion) with
respect to $I\in \left[ 0,S\left( P\right) +S\left( Q\right) \right] $.

Note the two limiting cases: when mutual information $I$ is zero
(corresponding to random matching), then $\mathcal{R}_{c}\left( 0\right)
=\left\{ C_{\infty }\right\} $, where $C_{\infty }^{k}=E_{p\otimes q}\left[
\phi ^{k}\left( X,Y\right) \right] $. When $I=S\left( P\right) +S\left(
Q\right) $, $\mathcal{R}_{c}\left( S\left( P\right) +S\left( Q\right)
\right) $ consists of the extreme points of the covariogram $\mathcal{F}_{c}$%
.

The following result combines the linearity embodied in (SLOI) and the
convex structure of the problem:

\begin{proposition}
\label{prop-cases}Under (O), (S), (GUI) and (SLOI),

a) The social welfare function $\mathcal{W}$ is positive homogeneous of
degree one in $\theta =\left( \Lambda ,\sigma \right) $. It is convex on $%
\mathbb{R}^{K}\times \lbrack 0,+\infty )$ and strictly convex on its
interior.

b) The subdifferential of $\mathcal{W}$ at $\left( \Lambda ,\sigma \right) $
is given by the set of $(K+1)$-uples%
\begin{equation*}
\partial \mathcal{W=}\left\{ \left( C(\Pi ),-I\left( \Pi \right) \right)
\right\}
\end{equation*}%
generated by an optimal matching $\Pi $ when parameter values $\theta
=\left( \Lambda ,\sigma \right) $ vary.

In particular, when the optimal matching $\Pi $ is unique for some $\theta $%
, then $\mathcal{W}$ is differentiable at $\theta $ and%
\begin{equation*}
C^{k}(\Pi )=\frac{\partial \mathcal{W}}{\partial \Lambda _{k}}(\theta
),~~~I\left( \Pi \right) =-\frac{\partial \mathcal{W}}{\partial \sigma }%
(\theta ),
\end{equation*}%
in which case we define
\begin{equation*}
C^{k}\left( \theta \right) :=\frac{\partial \mathcal{W}}{\partial \Lambda
_{k}}(\theta )\text{, and }I\left( \theta \right) :=-\frac{\partial \mathcal{%
W}}{\partial \sigma }(\theta ).
\end{equation*}

c) The function $I_{r}\left( C\right) $ is $C^{1}$ on the interior of $%
\mathcal{F}_{c}$, and one has
\begin{equation*}
\frac{\partial I_{r}}{\partial C^{k}}=\frac{\Lambda _{k}}{\sigma }.
\end{equation*}
\end{proposition}

As a corollary, the limiting homogeneous case also has interesting
comparative statics, which closely parallel the results above. When $\sigma
=0$ the mutual information does not play a role anymore in Theorem~\ref%
{theorem-socialw-primal}; so we focus on the covariogram $\mathcal{F}_{C}$;
and we define $\mathcal{W}_{0}(\Lambda )=\mathcal{W}(\Lambda ,0)$. Note in
particular that the boundary of the covariogram---which is the polygon in
Figure~\ref{Figure:CovarD}---consists of the set of the covariation vectors $%
C_{0}(\Lambda )$ when $\Lambda $ varies.

\begin{corollary}[Homogeneous comparative statics]
\label{prop-homo-comp}Under (O), (S), (GUI) and (SLOI),

a) The function $\mathcal{W}_{0}$ is convex and positive homogeneous of
degree one in $\Lambda $.

b) The subdifferential of $\mathcal{W}_{0}$ at $\Lambda $ is given by the
set of $K$-uples%
\begin{equation*}
\partial \mathcal{W}_{0}=\left\{ C(\Pi )\right\}
\end{equation*}%
generated by an optimal matching $\Pi $ when $\Lambda $ varies and $\sigma
=0 $.

In particular, when the solution $\Pi $ is unique for some $\Lambda $, then $%
\mathcal{W}_{0}$ is differentiable at $\Lambda $ and%
\begin{equation*}
\frac{\partial \mathcal{W}_{0}}{\partial \Lambda _{k}}(\Lambda )=C_{0}(\Pi ).
\end{equation*}
\end{corollary}

Our basic result is that any vector of covariations $C$ that is feasible
(that belongs to $\mathcal{F}_{C}$) can be rationalized for a well-chosen
value of total heterogeneity. This is a byproduct of the following result,
which sums up the relationships between the sets we introduced:

\begin{proposition}
\label{prop-geometric}Under (O), (S), (GUI) and (SLOI),

a) The sets $\mathcal{R}_{c}\left( I\right) $ are the set of extreme points
of nested closed and convex sets that expand from $\left\{ C_{\infty
}\right\} $ to $\mathcal{F}_{c}$ as mutual information $I$ goes from 0 to $%
\overline{I}_{0}$.

b) Any point ${\hat{C}}\in \mathcal{F}_{c}$ belongs to exactly one frontier $%
\mathcal{R}_{c}\left( I\right) $, associated to the mutual information $%
I=I_{r}\left( {\hat{C}}\right) $.

c) For a point $C$ such that $I_{r}\left( C\right) $ is smooth, letting $%
\Lambda _{k}=\frac{\partial I_{r}\left( C\right) }{\partial C^{k}}$; then
along $\partial \mathcal{R}_{c}\left( I\right) $
\begin{equation}
\frac{dC^{i}}{dC^{j}}=-\frac{\Lambda _{j}}{\Lambda _{i}}  \label{TMS}
\end{equation}
\end{proposition}

\bigskip

Proposition~\ref{prop-geometric} is illustrated on figure \ref{Figure:CovarD}%
. Note that when we fix $\Lambda $ and increase $\sigma $ from 0 to $+\infty
$, the summary vector $(C,I)$ for the optimal matching moves continuously
from $(C_{0}(\Lambda ),I_{0}(\Lambda ))$ to $(C_{\infty },0)$; thus part a)
tells us that increasing $\sigma $ for given $\Lambda $ moves us from a
point on the boundary of $\mathcal{F}_{C}$ to $C_{\infty }$.

Proposition~\ref{prop-geometric} may come as a surprise to the reader: we
have imposed quite a few assumptions on the way, and yet it seems that our
model still cannot rule out any feasible covariation of types across
partners! (Observing a $\hat{C}$ that is outside of $\mathcal{F}$ is
impossible by construction.) Proposition~\ref{prop-geometric} tells us that
observing $\hat{C}$ in the interior of $\mathcal{F}_C$ rejects the
homogeneous model; but that any such $\hat{C}$ can be rationalized by adding
the right amount of unobserved heterogeneity.

The interpretation of part c) is simplest when the matrix $\Lambda $ is
diagonal. With several dimensions for types, the optimal matching must
sacrifice some covariation in one dimension to the benefit of some
covariation in another. The implied sacrifice ratio, quite naturally, is
exactly the ratio of the assorting weights along these dimensions. Take for
instance the homogeneous case with only two characteristics, and set $%
\Lambda _{11}=1$ and $\Lambda _{22}=\varepsilon $. Then the function $%
\varepsilon \rightarrow C^{11}\left( 1,\varepsilon \right) $ is decreasing,
and the function $\varepsilon \rightarrow C^{22}\left( 1,\varepsilon \right)
$ is increasing. Therefore, when one puts more weight on the second
dimension, the covariation of the characteristics in the second dimension
increases, while the covariation on the first dimension decreases. Quite
intuitively, in the limit where all the weights are put on one dimension,
the classical Beckerian theory of positive assortative matching obtains.

More precisely, \citep{CGS:08} have shown in an $r$-type homogeneous
model that when $\Lambda _{11}=1$ and $\Lambda _{jj}\rightarrow 0$ for $%
j\geq 2$, if $\Pi ^{\ast }(\Lambda )$ is the $\Lambda $-optimal matching,
and $\left( X,Y\right) \sim \Pi ^{\ast }(\Lambda )$, then the joint
distribution of the first characteristics $\left( X^{1},Y^{1}\right) $
converges towards the maximally correlated distribution. Equivalently, $%
X^{1} $ and $Y^{1}$ become comonotonic in the limit, just as in classical
positive assortative matching.

\subsection{Uniqueness and purity\label{smoothsection}}

\textbf{Uniqueness. }As mentioned earlier, the boundary of $\mathcal{F}_{c}$
has kinks when types are discrete. In the homogeneous model ($\sigma =0$),
the optimal matching is pure for almost all values of $\Lambda $. Start from
such a value $\Lambda _{0}$. A small change in the value of $\Lambda $ will
not change the optimal matching $\Pi _{0}$, or the covariations it
generates. Pick some direction in $\Lambda $-space and move further away for
$\Lambda _{0}$. At some point $\Lambda _{1}$, the optimal matching will
change to a different pure matching, say $\Pi _{1}$; but this new pure
matching will vary with the direction we used to move away from $\Lambda
_{0} $. This is what generates kinks. Note also that in $\Lambda _{1}$, any
matching that is a convex combination of $\Pi _{0}$ and of $\Pi _{1}$ is
also optimal. So kinks are related to non-uniqueness of the optimal
matching. More formally:

\begin{proposition}
Assume (DD): the distributions $P$ and $Q$ are discrete. Then feasible set $%
\mathcal{F}_{c}$ is a polytope with a finite number of vertices that
correspond to pure matchings.
\end{proposition}

When there is enough observed heterogeneity, the optimal matching is unique.
Indeed, \citep{DMS:09} have shown that when the total heterogeneity $%
\sigma $ is large enough (so that $I$ is small enough), the solution to Eq. (%
\ref{schrodinger-eq}) is unique.

\bigskip

\textbf{Purity.} A matching is pure if a given type of man cannot be matched
to more than one type of women and conversely. Intuition suggests that given
sufficient heterogeneity, the optimal matching will not be pure, and its
probability weights will react to even small changes in $\Lambda $. In fact,
we have an even stronger result: even tiny levels of heterogeneity will make
the optimal matching impure. To see this, reason by contradiction: take a $%
\sigma >0$ and any given $\Lambda $ for which the optimal matching $\Pi $ is
pure. The objective function is%
\begin{equation*}
\sum_{x,y}\left( \pi (x,y)\;\Lambda \cdot \phi ^{k}(x,y)-\sigma \pi
(x,y)\log \frac{\pi (x,y)}{p(x)q(y)}\right) .
\end{equation*}%
Note that the derivative with respect to any $\pi (x,y)$ is infinite in $\pi
(x,y)=0$ and is finite anywhere else. Since $\Pi $ is pure, for any $x$
there is only one $y$ for which $\pi (x,y)$ is nonzero. Subtract a positive $%
\varepsilon $ from each such $\pi (x,y)$, and spread it over all zero
elements. The new joint distribution is still a feasible matching, and the
gain in social surplus from formerly zero probabilities outweighs the loss
from other matches. Therefore a pure matching cannot be optimal.

\section{Identification\label{sec:ident}}

The results in the previous sections give a very useful description of the
optimal matchings, and they show that $\sigma _{1}$ and $\sigma _{2}$ cannot
be identified separately. On the other hand, we have not provided a proof of
identification of the remaining parameters yet. We now set out to do so make
use for identification purposes of the geometrical interpretation of the
matching problem when the observable surplus is a linear combination of
known basis functions---this is assumption (SLOI), which we impose
throughout this section.

\subsection{Nonparametric idenfication}

Remember that given assumptions (O) and (S), there exist two functions $%
U(x,y)+V(x,y)=\Phi (x,y)$ such that the optimal matching obtains when man $%
\tilde{x}$ maximizes $U(x,y)+\chi (\tilde{x},y)$ over $y$ and woman $\tilde{x%
}$ maximizes $V(x,y)+\xi (\tilde{y},x)$\ over $x$. Now if $\pi $ is the
observable component of an optimal matching, it was showed in Section \ref%
{sec:optimal} that given assumption (GUI),
\begin{eqnarray*}
U(x,y) &=&\sigma _{1}\log \pi (x,y)+\sigma _{1}\log \frac{n_{1}(x)}{p(x)};%
\text{ and similarly,} \\
V(x,y) &=&\sigma _{2}\log \pi (x,y)+\sigma _{2}\log \frac{n_{2}(y)}{q(y)}.
\end{eqnarray*}

Now $U$ and $V$ depend on $\theta $ and are not easy to characterize as we
will see; but we know that they sum up to $\Phi $, so that
\begin{equation*}
\Phi (x,y)=\sigma \log \pi (x,y)+\sigma _{1}\left( \log n_{1}(x)-\log
p(x)\right) +\sigma _{2}\left( \log n_{2}(y)-\log q(y)\right) .
\end{equation*}%
In this formula $n_{1}$ and $n_{2}$ still depend on $\theta $ in a complex
way; but they only appear in terms that depend only on characteristics of
one partner. This means that the surplus function $\Phi $ is identified up
to an additive function of the form $a\left( x\right) +b\left( y\right) $.

To state this more formally, define the \emph{cross-difference operator\/}
as
\begin{equation*}
\Delta _{2}F(x,y;x^{\prime },y^{\prime })=\left( F(x^{\prime },y^{\prime
})-F(x^{\prime },y)\right) -\left( F(x^{\prime },y)-F(x,y)\right) ,
\end{equation*}%
for any function $F$ of $(x,y)$. Then we have:

\begin{theorem}[Cross-differences are identified up to scale]
\label{thm-crossders-ident} Assume (O), (S), (GUI) and (DD). For $\theta
=(\Lambda ,\sigma _{1},\sigma _{2})$ with $\sigma =\sigma _{1}+\sigma _{2}>0$%
, one has:

(i) There exists a unique optimal observable matching $\pi $ which maximizes
the social welfare (\ref{DualPT}).

(ii) There exist three vectors $\pi \left( x,y\right) $, $u(x)$ and $v(y)$,
and a constant $c$ normalized by $E_{p}\left[ u\left( X\right) \right] =E_{q}%
\left[ v\left( Y\right) \right] =0$, which are unique solutions to the
following system
\begin{equation}
\left\{
\begin{array}{c}
\pi \left( x,y\right) =p\left( x\right) q\left( y\right) \exp \left( \frac{%
\Phi (x,y)-u(x)-v(y)-c}{\sigma }\right) , \\
\pi \in \mathcal{M}\left( P,Q\right) .%
\end{array}%
\right.  \label{schrodinger-eq}
\end{equation}%
Further, the constant $c$ so defined coincides with the value of the social
welfare $c=\mathcal{W}$.

(iii) The probability $\pi $ defined in (ii) coincides with the optimal
matching solution of (\ref{DualPT}).
\end{theorem}

This result expresses that by adjusting the functions $u$ and $v$ at the
right level, one manages to satisfy the \textquotedblleft budget
constraint\textquotedblright\ that the matching has the right marginals
distributions $\pi \in \in \mathcal{M}\left( P,Q\right) $: hence, these
functions $u$ and $v$ can be interpreted as \textquotedblleft shadow
prices\textquotedblright\ of men and women's observable characteristics.

Theorem~\ref{thm-crossders-ident} has another consequence: the
complementarity of dimensions $i$ of the observable types of the partners in
$(x,y)$ can be tested directly on $\log \pi $, since $\Delta _{2}\log \pi $
and $\Delta _{2}\Phi $ have the same sign. Moreover, the relative strengths
of complementarities along dimensions $i$ and $j$ at a point $(x,y)$ can be
estimated by evaluating $\Delta _{2}\log \pi $ for values of $(x^{\prime
},y^{\prime })$ that differ from $(x,y)$ along these dimensions.

Theorem~\ref{thm-crossders-ident} immediately gives us an estimator of the
observable joint surplus function $\Phi$, up to additive functions of $x$
and of $y$. But adding any combination $a(x)+b(y)$ to the joint surplus does
not change the optimal matching, as long as we are determined not to have
singles---as we assume throughout the paper; and the positive scale factor $%
\sigma$ is irrelevant. So for instance $\log\hat{\pi}$ is a perfectly good
estimator of $\Phi$ if $\hat{\pi}$ consistently estimates $\pi$.

When we add a parametric structure under (SLOI), Theorem~\ref%
{thm-crossders-ident} also gives us an estimator of the assorting weights $%
\Lambda $ and the total heterogeneity $\sigma $\footnote{%
Recall that $\sigma _{1}$ and $\sigma _{2}$ are not separately identified.}.
In fact, the cross-difference operator is linear and so under (SLOI),
\begin{equation*}
\Delta_2 \log \pi=\frac{\Delta_2\Phi}{\sigma}=\sum_{k=1}^K \frac{\Lambda_k}{%
\sigma} \Delta_2 \phi^k;
\end{equation*}
if the cross-differences of the $\phi^k$ are linearly independent, then
observing $\pi$ gives us the $\Lambda$'s (along with overidentifying
restrictions.) This is a very weak requirement; having linearly dependent
basis functions would indeed be a modelling mistake.

This can be very simple in practice; to illustrate, take the diagonal
version of the (ER) example. Then if in $(x_{1},y_{1})$ man and woman are
both dropouts, keeping their income classes unchanged and moving them to
graduate level in $(x_{1}^{\prime },y_{1}^{\prime })$ generates
\begin{equation*}
\Delta _{2}\Phi (x_{1},y_{1};x_{1}^{\prime },y_{1}^{\prime })=\Lambda
_{1}^{E}-\Lambda _{0}^{E}.
\end{equation*}%
On the other hand, taking man and woman to have different education levels
in $(x_{2},y_{2})$ and swapping their educations to create $(x_{2}^{\prime
},y_{2}^{\prime })$ (again keeping income classes fixed) generates
\begin{equation*}
\Delta _{2}\Phi (x_{2},y_{2};x_{2}^{\prime },y_{2}^{\prime })=-\Lambda
_{1}^{E}-\Lambda _{0}^{E}.
\end{equation*}%
Therefore we obtain for instance
\begin{equation*}
\frac{\Lambda _{1}^{E}}{\Lambda _{0}^{E}}=\frac{\Delta _{2}\log \pi
(x_{2},y_{2};x_{2}^{\prime },y_{2}^{\prime })-\Delta _{2}\log \pi
(x_{1},y_{1};x_{1}^{\prime },y_{1}^{\prime })}{\Delta _{2}\log \pi
(x_{1},y_{1};x_{1}^{\prime },y_{1}^{\prime })+\Delta _{2}\log \pi
(x_{2},y_{2};x_{2}^{\prime },y_{2}^{\prime })},
\end{equation*}%
which is readily estimated from the observed matching.

These results are reminiscent of those in \citep{Fox:identmatchinggames}%
, although we obtained them under quite a different set of assumptions: we
do not use variation across subpopulations, neither does his rank-order
condition apply to our model. Note also that when specialized to
one-dimensional types, our result yields that of \citep{Siow:testbecker}
on testing complementarity of the surplus function by examining
log-supermodularity of the match distribution.

\subsection{Parametric identification}

Our parametric identification strategy will be either based on the knowledge
of the matching summaries $\left( \hat{C},\hat{I}\right) $, which are the
sufficient statistics for our model, or of just the covariation $\hat{C}$,
with the assumption that $\left( \hat{C},\hat{I}\right) $ lies on the
efficient frontier, that is $\hat{I}=I_{r}\left( \hat{C}\right) $. In either
cases, positive homogeneity imposes the need for a normalization of the
parameter $\left( \hat{\Lambda},\hat{\sigma}\right) $.

\subsubsection{The normalization rule}

Once again, $\theta $ is only identified up to a positive scale factor. Take
(SLOI) for instance: $\Lambda $ was only used to specify the objective
function, and so it can be multiplied by any positive constant without any
side-effect. In particular, $\mathcal{W}\left( t\Lambda ,t\sigma \right) =t%
\mathcal{W}\left( \Lambda ,\sigma \right) $ for $t\geq 0$. Therefore is is
quite clear that $\left( \Lambda ,\sigma \right) $ cannot be identified
without fixing some normalisation. So we normalize $\left( \Lambda ,\sigma
\right) $ by the choice
\begin{equation}
\text{\textbf{Normalization convention}: }\sigma I\left( \Lambda ,\sigma
\right) =1,  \label{normConv}
\end{equation}%
where as we recall, $I\left( \Lambda ,\sigma \right) =-\frac{\partial
\mathcal{W}\left( \Lambda ,\sigma \right) }{\partial \sigma }$.

Our general approach will be to identify the parameter value $\left( \hat{%
\lambda},1\right) $, and then rescale
\begin{equation*}
\hat{\Lambda}=\frac{\hat{\lambda}}{I\left( \hat{\lambda},1\right) },~~~\hat{%
\sigma}=\frac{1}{I\left( \hat{\lambda},1\right) }.
\end{equation*}

\subsubsection{Identification of $\protect\sigma $}

Note that $I_{r}\left( C\left( \lambda ,1\right) \right) =I\left( \lambda
,1\right) $, therefore, because of the normalization convention, $\sigma $
is identified by%
\begin{equation}
\hat{\sigma}=\frac{1}{I_{r}\left( \hat{C}\right) }.  \label{ident-sigma}
\end{equation}

\subsubsection{Identification of $\Lambda $}

As described above, we look for identifying $\hat{\lambda}$ among the
parameters of the form $\left( \lambda ,1\right) $. Remember%
\begin{equation*}
I_{r}\left( C\right) =\sup_{\lambda }\left\{ \lambda \cdot C-\mathcal{W}%
\left( \lambda ,1\right) \right\}
\end{equation*}%
so by the enveloppe theorem, $\hat{\lambda}=\frac{\partial I_{r}\left( \hat{C%
}\right) }{\partial C}$ is such that $C\left( \hat{\lambda},1\right) =\frac{%
\partial \mathcal{W}\left( \hat{\lambda},1\right) }{\partial \lambda }$.
Hence, $\Lambda $ is identified by%
\begin{equation}
\hat{\Lambda}=\frac{1}{I_{r}\left( \hat{C}\right) }\frac{\partial
I_{r}\left( \hat{C}\right) }{\partial C}.  \label{ident-lambda}
\end{equation}

\subsection{Comparative statics}

We define the \emph{best additive projector} $\mathcal{P}h$ of a vector $%
h\left( x,y\right) $ as%
\begin{equation*}
\mathcal{P}h\left( x,y\right) =f\left( x\right) +g\left( y\right)
\end{equation*}%
where $f$ and $g$ minimize%
\begin{equation*}
E_{\pi }\left[ \left( h\left( X,Y\right) -E\left[ h\left( X,Y\right) \right]
-f\left( X\right) -g\left( Y\right) \right) ^{2}\right] .
\end{equation*}

We have immediately that $E_{P}\left[ f\left( X\right) \right] =0$ and $E_{Q}%
\left[ g\left( Y\right) \right] =0$, and introducing the \emph{residue} $%
\varepsilon $
\begin{equation*}
\varepsilon \left( X,Y\right) =h\left( X,Y\right) -E\left[ h\left(
X,Y\right) \right] -f\left( X\right) -g\left( Y\right)
\end{equation*}%
we get $E\left[ \varepsilon \left( X,Y\right) |X\right] =0$ and $E\left[
\varepsilon \left( X,Y\right) |Y\right] =0$. The decomposition%
\begin{equation*}
h\left( X,Y\right) =E\left[ h\left( X,Y\right) \right] +f\left( X\right)
+g\left( Y\right) +\varepsilon \left( X,Y\right)
\end{equation*}%
is the \emph{two-way ANOVA}\ decomposition of $h\left( X,Y\right) $. The
following proposition will be the fundamental tool for inference. It
expresses that the projection residue in the two-way ANOVA decomposition of $%
\phi ^{k}$ is the score function $\sigma \frac{\partial \log \pi }{\partial
\Lambda _{k}}$.

\begin{proposition}[Score function]
\label{prop-partialsT}Under (O), (S), (GUI), and (SLOI), the score function
is given by
\begin{equation*}
\frac{\partial \log \pi }{\partial \Lambda _{k}}(x,y)=\frac{\phi ^{k}\left(
x,y\right) -\mathcal{P}\phi ^{k}\left( x,y\right) -E\left[ \phi ^{k}\left(
X,Y\right) \right] }{\sigma },
\end{equation*}

that is $\frac{\partial u\left( x\right) }{\partial \Lambda _{k}}+\frac{%
\partial v\left( y\right) }{\partial \Lambda _{k}}=\mathcal{P}\phi
^{k}\left( x,y\right) $, where $u$ and $v$ are solution to Equation (\ref%
{schrodinger-eq}).
\end{proposition}

As a result, we get an expression for the computation of the Hessian of the
social welfare function at fixed $\sigma $.

\begin{proposition}[Fisher information matrix]
\label{hessian}Under (O), (S), (GUI), and (SLOI), wherever $\mathcal{W}%
\left( \Lambda ,\sigma \right) $ is derivable, we get%
\begin{equation*}
\frac{\partial ^{2}\mathcal{W}\left( \Lambda ,\sigma \right) }{\partial
\Lambda _{k}\partial \Lambda _{l}}=\sigma \mathcal{I}^{kl}\left( \theta
\right) ,
\end{equation*}%
where
\begin{equation*}
\mathcal{I}^{kl}\left( \theta \right) :=E\left[ \frac{\partial \log \pi }{%
\partial \Lambda _{k}}(X,Y)\frac{\partial \log \pi }{\partial \Lambda _{l}}%
(X,Y)\right]
\end{equation*}%
is the Fisher information matrix. Further,%
\begin{equation}
\mathcal{I}^{kl}\left( \theta \right) :=\frac{cov\left( \phi ^{k}\left(
X,Y\right) ,\phi ^{l}\left( X,Y\right) \right) -cov\left( \mathcal{P}\phi
^{k}\left( X,Y\right) ,\mathcal{P}\phi ^{l}\left( X,Y\right) \right) }{%
\sigma ^{2}}.  \label{Fisher}
\end{equation}
\end{proposition}

\section{Inference\label{sec:inference}}

We now turn to the problem of inference. Our data will consist of matched
characteristics of $N$ pairs $\left\{ \left( x_{1},y_{1}\right) ,...,\left(
x_{N},y_{N}\right) \right\} $, and our null hypothesis is that they were
generated by an optimal matching consistent with assumptions (O), (S),
(GUI), and (SLOI). Given a proposed specification for the basis functions $%
\phi ^{k}$, and our estimates of the marginal distributions of types $\hat{P}%
_{N}$ and $\hat{Q}_{N}$, we would therefore like to infer the values of $%
\Lambda $ and $\sigma $ which come closest to rationalizing the observed
matching. We use our theory to answer two questions:

\begin{enumerate}
\item is the observed matching optimal?

\item which parameter vector $\left( \Lambda ,\sigma \right) $ best
rationalizes the observed matching (exactly if the observed matching is
optimal, approximately if it is not)?
\end{enumerate}

\bigskip

The primary object of our investigation will be the empirical moments of $%
\phi ^{k}$,%
\begin{equation*}
\hat{C}_{N}^{k}=\frac{1}{N}\sum_{n=1}^{N}\phi ^{k}\left( x_{n},y_{n}\right) .
\end{equation*}

Let $C^{k}$ denote the expectation of $\phi ^{k}(X,Y)$ under the joint
distribution $\Pi $ of $(X,Y)$. Standard asymptotic theory of the empirical
process (\citep{Vaart:1998}) implies the convergence in distribution
\begin{equation*}
\sqrt{N}\left( \hat{C}_{N}^{k}-C^{k}\right) \Longrightarrow \xi ^{k}
\end{equation*}%
where $\xi ^{k}=\int \phi ^{k}\left( x,y\right) dG\left( x,y\right) $, $G$
being a $\Pi $-Brownian bridge. In particular,
\begin{equation*}
cov\left( \sqrt{N}\left( \hat{C}_{N}^{k}-C^{k}\right) ,\sqrt{N}\left( \hat{C}%
_{N}^{l}-C^{l}\right) \right) =cov_{\Pi }\left( \phi ^{k}\left( X,Y\right)
,\phi ^{l}\left( X,Y\right) \right)
\end{equation*}%
for all $1\leq k,l\leq K$.

We shall call $\mathcal{W}_{N}\left( \theta \right) $ the value of the
social surplus at parameter $\theta $ obtained with the empirical
distributions of observable types $P_{N}$ and $Q_{N}$.

\bigskip

\textbf{Normalization.} Recall that because of positive homogeneity, models $%
\theta =\left( \Lambda ,\sigma \right) $ and $t\theta =\left( t\Lambda
,t\sigma \right) $ are observationally indistinguishable. Just as in the
previous section, we impose the normalization convention $\sigma I\left(
\Lambda ,\sigma \right) =1$. When we describe estimators below, we first
compute an estimator of the assorting weights $\Lambda $ for total
heterogeneity $\sigma =1$; we denote it $\hat{\lambda}_{N}$. We then shall
get an estimator of the mutual information $\hat{I}_{N}$. To obtain the
normalized estimator in each case, the reader should divide the vector $(%
\hat{\lambda}_{N},1)$ by the scalar $\hat{I}$.

\bigskip

The results we obtained in sections~\ref{sec:optimal} and~\ref{sec:ident}
suggest two estimation strategies, which we will now define and compare.

\subsection{Nonparametric inference\label{est:semipar}}

Theorem~\ref{thm-crossders-ident} and its corollary immediately suggest a
very simple nonparametric approach. In this discrete case, a nonparametric
estimator $\hat{\pi}_{N}(x,y)$ is readily obtained, by counting the
proportion of matches between a man of characteristics $x$ and a woman of
characteristics $y$. We could pick arbitrary functions $a(x)$ and $b(y)$ and
define
\begin{equation*}
\hat{\Phi}_{N}(x,y)=\log \hat{\pi}_{N}(x,y)+a(x)+b(y),
\end{equation*}%
without any reference to basis functions---imposing $\sigma =1$ on the way.
Then if we further assume (SLOI) with basis functions $\phi ^{k}$, we can
apply minimum-distance techniques to recover an estimator $\hat{\lambda}%
_{N}^{SP}$, which minimizes some norm
\begin{equation*}
\Vert \hat{\Phi}_{N}(x,y)-\lambda \cdot \phi (x,y)\Vert .
\end{equation*}%
Note that as usual, the minimum value of the norm allows us to construct a
test statistic for the hypothesis that $\Phi $ is a linear combination of
the $\phi ^{k}$.

More generally, we know that under (O), (S) and (GUI) only,
\begin{equation*}
\log \pi =\frac{\Phi }{\sigma };
\end{equation*}%
thus a nonparametric estimate $\hat{\pi}_{N}$ can be used as a heuristic
device to decide on a set of basis functions, and/or to test for the
adequacy of such a set.

We now turn to parametric estimators.

\subsection{Parametric inference:\ The Moment Matching Estimator}

Our second estimator is based solely on the statistics of the matching
covariations $\hat{C}$. It rests on identification of $\Lambda $ provided by
Eq. (\ref{ident-lambda}). Therefore $\hat{\lambda}$ is taken as a maximizer
of
\begin{equation}
\Lambda \cdot \hat{C}-\mathcal{W}_{N}\left( \Lambda ,1\right)
\label{Legendre}
\end{equation}%
over all possible $\Lambda $. This being a strictly concave function, its
minimizer is unique; further efficient computation is available. Letting $%
\hat{I}$ the value of expression (\ref{Legendre}) at the optimal value $\hat{%
\lambda}$, we obtain the \textbf{Moment Matching (MM) estimator, denoted }$%
\hat{\Lambda}^{MM}$ and $\hat{\sigma}^{MM}$, by setting\textbf{\ }%
\begin{equation*}
\hat{\Lambda}^{MM}=\frac{\hat{\lambda}}{\hat{I}},~~~\hat{\sigma}^{MM}=\frac{1%
}{\hat{I}}.
\end{equation*}

Now if our data was generated by an optimal matching $\Pi $ for parameters $%
\left( \hat{\Lambda}^{MM},\hat{\sigma}^{MM}\right) $, the empirical
covariations $\hat{C}_{N}$ would coincide with the optimal correlations $%
C\left( \hat{\Lambda}^{MM},\hat{\sigma}^{MM}\right) $. By construction, the
MM estimator is the value of assorting weights $\lambda $ such that the
predicted covariations coincide with the observed covariations. The Moment
matching estimator is consistent and asymptotically Gaussian, and

\begin{theorem}
\label{thm-asymptotic}Under (O), (S), (GUI) and (SLOI),
\begin{equation*}
\sqrt{N}\left( \hat{\lambda}_{N}-\lambda \right) \Longrightarrow \mathcal{I}%
^{-1}\xi
\end{equation*}%
where $\xi $ is the Brownian bridge characterized at the beginning of this
section and the matrix $\mathcal{I}^{kl}$ is the Fisher information matrix
expressed above in (\ref{Fisher}). In particular, the MM estimator is
asymptotically efficient.
\end{theorem}

\section{Computational issues}

With the exception of the semiparametric estimator (SP), our inferential
methods require solving for the optimal matching for potentially large
populations, and a large number of parameter vectors during optimization.
This may seem to be a forbidding task: there exist well-known algorithms to
find an optimal matching, and they are reasonably fast; but with large
populations the required computer resources may still be large.

Fortunately, it turns out that introducing (our type of) heterogeneity
actually makes computing optimal matchings much simpler; this is a boon for
the ML and MM estimators\footnote{%
The BP estimator is designed for the homogeneous case and so the following
does not apply to it.}.

To see this, choose a parameter vector $\theta=(\Phi,\sigma)$ and return to
the characterization of optimal matchings in equation~\ref{DualPT}, in the
continuous case (CD) for simplicity. Dividing by $\sigma$ and taking the
logarithm, optimal matchings can also be obtained by solving the following
minimization program:
\begin{equation*}
\min_{\Pi \in \mathcal{M}(P,Q)} \sum_{x,y} \pi(x,y)\log \frac{\pi(x,y)}{%
p(x)q(y)\exp(\Phi(x,y)/\sigma)}.
\end{equation*}
Now define a set of probabilities $r$ by
\begin{equation*}
r(x,y)=\frac{p(x)q(y)\exp(\Phi(x,y)/\sigma)}{\sum_{x,y}
p(x)q(y)\exp(\Phi(x,y)/\sigma)};
\end{equation*}
and note that given any choice of parameters $\theta$ and known marginals $%
(p,q)$, the probability $r$ itself is known.

Determining the optimal matchings therefore boils down to finding the joint
probabilities $\pi$ with known marginals $p$ and $q$ which minimize the
Kullback-Leibler distance to $r$:
\begin{equation}
\sum_{x,y} \pi(x,y) \log \frac{\pi(x,y)}{r(x,y)}.  \label{eqIPFP}
\end{equation}
Equivalently, we are looking for the Kullback-Leibler projection of $r$ on $%
\mathcal{M}(P,Q)$.

This is a well-known problem in various fields, and algorithms to solve it
have been around for a long time. National accountants, for instance, use
RAS algorithms to fill cells of a two-dimensional table whose margins are
known; here the choice of $r$ reflects prior notions of the correlations of
the two dimensions of the table. These RAS algorithms belong to a family
called Iterative Projection Fitting Procedures (IPFP). They are very fast,
and are guaranteed to converge under weak conditions. We only describe the
application of IPFP to our model here; we direct the reader to %
\citep{Ruschendorf:95} for more information.

The intuition of equation~\ref{eqIPFP} is quite clear: the random matching,
which is optimal when $\sigma$ is very large, has $\pi(x,y)=p(x)q(y)$. For
smaller $\sigma$' s the probability of a match between $x$ and $y$ must
increase with the surplus it creates, $\Phi(x,y)$; and given our assumption
(GUI) on the distribution of unobserved heterogeneity, it should not come as
a surprise that the corresponding factor is multiplicative and exponential.

To describe the algorithm, we split $\pi$ into\footnote{%
It can be shown that at the optimum $\pi(x,y)=0$ where $r(x,y)=0$.}
\begin{equation*}
\pi(x,y)=r(x,y)\exp(-(u(x)+v(y))/\sigma).
\end{equation*}
The functions $u$ and $v$ of course will only be determined up to a common
constant. The algorithm iterates over values $(u^k, v^k)$. We start from $%
u^0 \equiv -\sigma\log p$ and $v^0 \equiv 0$. Then at step $(k+1)$ we
compute
\begin{equation*}
\exp(-v^{k+1}(y)/\sigma)=\frac{q(y)}{\sum_x r(x,y) \exp(-u^k(x)/\sigma) }
\end{equation*}
and
\begin{equation*}
\exp(-u^{k+1}(x)/\sigma)=\frac{p(x)}{\sum_y r(x,y) \exp(-v^{k+1}(y)/\sigma)}.
\end{equation*}
Two remarks are in order here: first, we could just as well start from $%
u^0\equiv 0$ and $v^0=-\sigma\log q$ and modify the iteration formul\ae\ %
accordingly. Second and just as in other Gauss-Seidel algorithms, it is
important to update one component based on the other updated component: the
right-hand sides have $u^k$ and $v^{k+1}$.

If $(u,v)$ is a fixed point of the algorithm, then
\begin{equation*}
\frac{\pi(x,y)}{p(x)q(y)}=\exp\left(\frac{\Phi(x,y)-u(x)-v(y)}{\sigma}%
\right).
\end{equation*}
Comparing this formula to Theorem~\ref{thm-crossders-ident} shows the
benefit of this reparameterization, since $u(x)$ and $v(y)$ have a simple
interpretation: they represent (up to a common additive constant) the
expected utilities of a man of observable characteristics $x$ and of a woman
of observable characteristics $y$. This can be seen by checking, for
instance, that
\begin{equation*}
E(\max U(X,Y)|X=x)=\sigma_1 \log n_1(x).
\end{equation*}
Thus the IPFP algorithm gives us not only the optimal matching, but also
these expected utilities.

The simplification does not stop there. In fact, given data on $N$ couples,
the marginal $p$ assigns $1/N$ probability to each of $(x_{1},\ldots ,x_{N})$%
, and similarly for women. Define a matrix $\Psi $ by $\Psi _{ij}=\exp (\Phi
(x_{i},y_{j})/\sigma )$, and vectors $a_{i}^{k}=\exp (-u^{k}(x_{i})/\sigma )$%
, $b_{j}^{k}=\exp (-v^{k}(y_{j})/\sigma )$. Then we end up with the
shockingly simple and inexpensive formul\ae :
\begin{equation*}
b^{k+1}=\frac{N}{\Psi ^{\prime }a^{k}}\;\mbox{ and }\;a^{k+1}=\frac{N}{\Psi
b^{k+1}}.
\end{equation*}

\section{Possible extensions and concluding remarks\label{sect-extensions}}

Our theory so far relies on several strong assumptions. Some of them are
easy to relax; we discuss three of them, before turning to potential
extensions.

\bigskip

\textbf{Single households.} So far we have not allowed for unmatched
individuals. In an optimal matching, some men and/or women may remain
single, as of course some must if there are more individuals on one side of
the market. The choice of the socially optimal matching can be broken down
into the choice of the set of individuals who participate in matches and the
choice of actual matches between the selected men and women. Our theory
applies without any change to the second subproblem; that is, all of our
results extend to $M$ and $W$ as selected in the first subproblem.

From the point of view of statistical inference, we may lose some efficiency
in doing so; we note here that when the unobserved heterogeneity in
preferences over partners is separable from the utility of marriage itself,
our method does not incur any efficiency loss.

\bigskip

\textbf{Non-bipartite matching.} Bipartite matching refers to the fact that
each individual is exogenously assigned in one category---in our
terminology, husband or wife. Our analysis in fact is very easy to extend so
as to incorporate same-sex unions, and thus to rationalize endogamy in the
gender dimension.

To do so, we just need to add one (observed) characteristic, in the form of
gender. If for instance gender becomes the first dimension of the
characteristics vector, then the observed surplus has an assorting weight $%
\Lambda _{11}<0$ that reflects the more typical preference for the opposite
sex; while heterogenous preferences $\chi $ and $\eta $ will automatically
take into account the dispersion of individual preference for same-sex
unions.

\bigskip

\textbf{Continuous distributions.} While we have assumed discrete
characteristics, we expect the main thrust of our arguments to carry over to
the case where the distributions of the characteristics are continuous. We
are working on such an extension; this will require adapting the (GUI)
assumption to one that is better-suited to continuous choice.

\bigskip

\textbf{Revealed Preferences.} As mentioned in the section on the Boundary
Projection estimator, the Lagrange multiplier $e$ is known in the theory of
revealed preferences as Afriat's efficiency index. The analogy in fact goes
deeper. Recall the basic theorem on revealed preferences:

\begin{proposition}[Afriat]
The following conditions are equivalent:

(i) The observed quantity-price vectors $\left( x_{k},p _{k}\right)_{ k=1}^N
$ are consistent with maximization of a single utility function;

(ii) There exist scalars $\lambda _{k}<0,$ $k=1,...,N$ such that
\begin{equation*}
\sum_{k=1}^{N}\lambda _{k} \; p_k \cdot x_{\sigma \left( k\right) }
\end{equation*}%
is maximized over $\sigma \in \mathfrak{S}_{N}$ when $\sigma(j) =j$ for all $%
j$.
\end{proposition}

This is reminiscent of a multidimensional matching problem in which prices $%
p $ correspond to the characteristics $x$ of men, consumptions $q$ to those
of women $y$, and there is no unobserved heterogeneity. We are currently
exploring this nalogy.

\bigskip

\textbf{Screening.} In the theory of screening, a \textquotedblleft
type\textquotedblright\ $\theta $ refers to a set of individual
characteristics that are privately observed. Assume that utilities are
additively separable in transfers, with
\begin{equation*}
u(q,\theta )-t\;\mbox{ for an agent of type }\;\theta
\end{equation*}%
and
\begin{equation*}
W(q)+t\;\mbox{ for the principal}.
\end{equation*}%
Then given quantity-transfer pairs $(q_{k},t_{k})_{k=1}^{N}$ that presumably
correspond to different types, it can be shown that
\begin{equation*}
\sum_{k=1}^{N}u(q_{k},\theta _{\sigma (k)})
\end{equation*}%
is maximized over $\sigma \in \mathfrak{S}_{N}$ when $\sigma (j)=j$ for all $%
j$.

This again suggests that our methods may help in estimating screening models.

\newpage

\appendix

\section{Facts from Convex Analysis}

\subsection{Basic results\label{app-Convexity}}

We only sum up here the concepts we actually use in the paper; we refer the
reader to \citep{HL:01} for a thorough exposition of the topic.

Take any set $Y\subset \mathrm{I\kern-.17emR}^{d}$; then the \emph{convex
hull\/} of $Y$ is the set of points in $\mathrm{I\kern-.17emR}^{d}$ that are
convex combinations of points in $Y$. We usually focus on its closure, the
closed convex hull, denoted $cch\left( Y\right) $.

The \emph{support function} $S_{Y}$ of $Y$ is defined as
\begin{equation*}
S_{Y}\left( x\right) =\sup_{y\in Y}x\cdot y
\end{equation*}%
for any $x$ in $Y$. It is a convex function, and it is homogeneous of degree
one. Moreover, $S_{Y}=S_{\mbox{cch}\left( Y\right) }$ where $\mbox{cch}%
\left( Y\right) $ is the closed convex hull of $Y$, and $\partial
S_{Y}\left( 0\right) =\mbox{cch}\left( Y\right) $.

A point in $Y$ is an \emph{extreme point} if it does not belong in any open
line segment joining two points of $Y$.

Now let $u$ be a convex, continuous function defined on $\mathrm{I\kern%
-.17emR}^{d}$. Then the gradient $\nabla u$ of $u$ is well-defined almost
everywhere and locally bounded. If $u$ is differentiable at $x$, then
\begin{equation*}
u\left( x^{\prime }\right) \geq u\left( x\right) +\nabla u\left( x\right)
\cdot (x^{\prime }-x)
\end{equation*}%
for all $x^{\prime }\in \mathrm{I\kern-.17emR}^{d}$. Moreover, if $u$ is
also differentiable at $x^{\prime }$, then
\begin{equation*}
\left( \nabla u\left( x\right) -\nabla u\left( x^{\prime }\right) \right)
\cdot \left( x-x^{\prime }\right) \geq 0.
\end{equation*}%
When $u$ is not differentiable in $x$, it is still \emph{\
subdifferentiable\/} in the following sense. We define $\partial u\left(
x\right) $ as%
\begin{equation*}
\partial u\left( x\right) =\left\{ y\in \mathrm{I\kern-.17emR}^{d}:\forall
x^{\prime }\in \mathrm{I\kern-.17emR}^{d},u\left( x^{\prime }\right) \geq
u\left( x\right) +y\cdot (x^{\prime }-x)\right\} .
\end{equation*}%
Then $\partial u\left( x\right) $ is not empty, and it reduces to a single
element if and only if $u$ is differentiable at $x$; in that case $\partial
u\left( x\right) =\left\{ \nabla u\left( x\right) \right\} $.

\subsection{Generalized Convexity\label{app-GalConvexity}}

In order to make the paper self-contained, we present basic results on the
theory of \textit{generalized convexity}, sometimes called the theory of $c$%
-convex functions. This theory extends many results from convex analysis
and, in particular, duality results, \ to a much more general setting. We
refer to \citep{Villani:2009}, p. 54--57 (or \citep{Villani:2003},
pp. 86--87) for a detailed account\footnote{%
A cautionary remark is in order here: the sign conventions vary in the
literature, so our own choices may differ from those of any given author.}.

Let $\omega$ be a function from the product of two sets $\mathcal{X}\times
\mathcal{Y}$ to $\lbrack -\infty ,+\infty )$.

\begin{definition}
\label{def:galconvexity} Consider any function $\psi :\mathcal{X}\rightarrow
(-\infty ,+\infty ]$. Its generalized Legendre transform $\psi ^{\bot }:%
\mathcal{X}\rightarrow \lbrack -\infty ,+\infty )$ is defined by%
\begin{equation*}
\psi ^{\bot }\left( y\right) =\inf_{x\in \mathcal{X}}\left\{ \psi \left(
x\right) -\omega \left( x,y\right) \right\} .
\end{equation*}

Conversely, take any function $\zeta :\mathcal{Y}\rightarrow \lbrack -\infty
,+\infty )$; then its generalized Legendre transform $\zeta ^{\top }:%
\mathcal{X}\rightarrow (-\infty ,+\infty ]$ is defined by%
\begin{equation*}
\zeta ^{\top }\left( x\right) =\sup_{y\in \mathcal{Y}}\left\{ \zeta \left(
y\right) +\omega \left( x,y\right) \right\} .
\end{equation*}

A function $\psi $ is called $\omega $-convex if it is not identically $%
+\infty $ and if there exists $\zeta :\mathcal{Y}\rightarrow \lbrack -\infty
,+\infty ]$ such that
\begin{equation*}
\psi =\zeta ^{\top }.
\end{equation*}
\end{definition}

Recall that the usual Legendre transform is defined as
\begin{equation*}
\psi^* (y)=\inf_{x \in \mathcal{X}}\left\{ \psi \left( x\right) - x \cdot y
\right\};
\end{equation*}
thus it coincides with the generalized Legendre transform when $\omega$ is
bilinear, and then $\omega$-convexity boils down to standard convexity.

\bigskip

Our analysis rests on the following fundamental result, which generalizes
standard convex analysis.

\begin{proposition}
\label{prop-top-bot} For every function $\psi :\mathcal{X}\rightarrow
(-\infty ,+\infty ]$,
\begin{equation*}
\psi ^{\bot \top }\leq \psi
\end{equation*}%
with equality if and only if $\psi $ is $\omega $-convex.
\end{proposition}

\begin{varproof}
Take any $x\in \mathcal{X}$; then
\begin{equation*}
\psi ^{\bot \top }\left( x\right) =\sup_{y\in \mathcal{Y}}\inf_{x^{\prime
}\in \mathcal{X}}\left\{ \psi \left( x^{\prime }\right) -\omega \left(
x^{\prime },y\right) +\omega \left( x,y\right) \right\};
\end{equation*}
taking $x^{\prime}=x$ shows that $\psi ^{\bot \top }\left( x\right)\leq\psi
\left( x\right)$.

Conversely, if $\psi^ {\bot \top}=\psi$ then $\psi \left( x\right) =\zeta
^{\top }\left( x\right) $, with $\zeta =\psi ^{\bot }$. But for any function
$\zeta$, the triple transform $\zeta ^{\top \bot \top}$ coincides with $%
\zeta^\top$. To see this, write
\begin{equation*}
\zeta ^{\top \bot \top }\left( x\right) =\sup_{y\in \mathcal{Y}}
\inf_{x^{\prime}\in \mathcal{X}} \sup_{y^{\prime }\in \mathcal{Y}}\left\{
\zeta \left( y^{\prime }\right) +\omega \left( x^{\prime },y^{\prime
}\right) -\omega \left( x^{\prime },y\right) +\omega \left( x,y\right)
\right\}.
\end{equation*}%
Now for all $x$ and $y$,
\begin{equation*}
\inf_{x^{\prime }\in \mathcal{X}}\sup_{y^{\prime }\in \mathcal{Y}}\left\{
\zeta \left( y^{\prime }\right) +\omega \left( x^{\prime },y^{\prime
}\right) -\omega \left( x^{\prime },y\right) \right\} \geq \zeta \left(
y\right)
\end{equation*}%
as is easily seen by taking $y^{\prime}=y$; therefore
\begin{equation*}
\zeta ^{\top \bot \top }\left( x\right) \geq \sup_{y\in \mathcal{Y}} \left\{
\zeta \left( y\right) +\omega \left( x,y\right)\right\}=\zeta^ \top(x).
\end{equation*}%
Applying this to the $\zeta$ such that $\psi=\zeta^\top$ concludes the proof.

QED.
\end{varproof}

\section{\label{app-Proofs}Proofs}

\subsection{Proof of Theorem \protect\ref{theorem-socialw-primalp}}

In order to prove Theorem 1, some preparation is needed. Remember our
shorthand notation $\tilde{x}=(x,\varepsilon )$, and $\tilde{y}=(y,\eta )$.
For any function $\tilde{u}\left( x,\varepsilon \right) $, fix $x$ and use
the theory of generalized convexity briefly recalled in Appendix (\ref%
{app-GalConvexity}) to define
\begin{equation*}
\tilde{u}^{\bot }\left( x,y\right) =\inf_{\varepsilon }\left\{ \tilde{u}%
\left( x,\varepsilon \right) -\chi \left( \left( x,\varepsilon \right)
,y\right) \right\}
\end{equation*}%
the \textit{generalized Legendre transform} of $\tilde{u}\left( x,\cdot
\right) $ with respect to the partial surplus function $\chi \left( \left(
x,\cdot \right) ,\cdot \right) $. We define in the same manner
\begin{equation*}
\tilde{v}^{\bot }\left( x,y\right) =\inf_{\eta }\left\{ \tilde{v}\left(
y,\eta \right) -\xi \left( x,\left( y,\eta \right) \right) \right\} .
\end{equation*}

Similarly, for two functions $U\left( x,y\right) $ and $V\left( x,y\right) $%
, we define%
\begin{eqnarray*}
U^{\top }\left( x,\varepsilon \right) &:&=\sup_{y}\left\{ U\left( x,y\right)
+\chi \left( \left( x,\varepsilon \right) ,y\right) \right\} \\
V^{\top }\left( y,\eta \right) &:&=\sup_{x}\left\{ V\left( x,y\right) +\xi
\left( x,\left( y,\eta \right) \right) \right\} .
\end{eqnarray*}

\begin{lemma}
\label{lemma:abstract-duality} Let $A$ be the set of pairs of functions $%
(U,V)$ such that
\begin{equation*}
\forall x,y, \; U\left( x,y\right) +V\left( x,y\right) \geq \Phi \left(
x,y\right).
\end{equation*}
Then
\begin{equation*}
\mathcal{W}=\inf_{ (U,V) \in A}\left\{ \int U^{\top }\left( \tilde{x}\right)
d\tilde{P}\left( \tilde{x}\right) +\int V^{\top }\left( \tilde{y}\right) d%
\tilde{Q}\left( \tilde{y}\right)\right\}.
\end{equation*}
\end{lemma}

\begin{varproof}[Proof of Lemma \protect\ref{lemma:abstract-duality}]
By the Kantorovich duality theorem (\citep{Villani:2009} Theorem 5.10),%
\begin{equation}
\mathcal{W}=\sup_{\tilde{\pi}\in \mathcal{M}\left( P,Q\right) }\int \tilde{%
\Phi}\left( \tilde{x},\tilde{y}\right) d\pi \left( \tilde{x},\tilde{y}%
\right) =\inf_{(\tilde{u},\tilde{v})\in \tilde{A}}\left\{ \int \tilde{u}%
\left( \tilde{x}\right) d\tilde{P}\left( \tilde{x}\right) +\int \tilde{v}%
\left( \tilde{y}\right) d\tilde{Q}\left( \tilde{y}\right) \right\} ,
\label{dualInitial}
\end{equation}%
where $\tilde{A}$ is the set of pairs of functions $(\tilde{u}.\tilde{v})$
such that
\begin{equation*}
\forall \tilde{x},\tilde{y},\;\tilde{u}\left( \tilde{x}\right) +\tilde{v}%
\left( \tilde{y}\right) \geq \tilde{\Phi}\left( \tilde{x},\tilde{y}\right) .
\end{equation*}%
Note the following two facts about the right-hand side of this equality:

\begin{enumerate}
\item Since
\begin{equation*}
\tilde{\Phi}(\tilde{x},\tilde{y})=\Phi(x,y)+\chi \left( \left( x,\varepsilon
\right) ,y\right) +\xi \left( \left( y,\eta \right) ,x\right),
\end{equation*}
the infimum in (\ref{dualInitial}) can be taken over the pair of functions $%
\left( \tilde{u},\tilde{v}\right) $ that satisfy
\begin{equation*}
\tilde{u}\left( x,\varepsilon \right) \geq \sup_{y}\left\{ \Phi (x,y)+\chi
\left( \left( x,\varepsilon \right) ,y\right) + \sup_{\eta }\left[ \xi
\left( \left( y,\eta \right) ,x\right) -\tilde{v}\left( y,\eta \right) %
\right] \right\},
\end{equation*}%
or
\begin{equation*}
\tilde{u}(\tilde{x}) \geq \sup_y \left\{ \Phi(x,y)+\chi((x,\varepsilon),y)-%
\tilde{v}^{\bot}(x,y). \right\}
\end{equation*}
At the optimum this must hold with equality. Going back to Definition~\ref%
{def:galconvexity}, it follows that $\tilde{u}\left( x,\cdot \right) $ is $%
\chi((x,\cdot),\cdot)$-convex for every $x$; and using Proposition~\ref%
{prop-top-bot}, we can substitute $\tilde{u}$ with $\tilde{u}^{\bot \top}$,
that is:
\begin{equation*}
\tilde{u}\left( x,\varepsilon \right) =\sup_{y}\left\{ \tilde{u}^{\bot
}\left( x,y\right) +\chi \left( \left( x,\varepsilon \right) ,y\right)
\right\}.
\end{equation*}

Given a similar argument on $\tilde{v}$, the objective function can be
rewritten as%
\begin{equation*}
\int \sup_{y}\left\{ \tilde{u}^{\bot }\left( x,y\right) +\chi \left( \left(
x,\varepsilon \right) ,y\right) \right\} d\tilde{P}\left( \tilde{x}\right)
+\int \sup_{x}\left\{ \tilde{v}^{\bot }\left( x,y\right) +\xi \left(
x,\left( y,\eta \right) \right) \right\} d\tilde{Q}\left( \tilde{y}\right).
\end{equation*}

\item Also note that the constraint of the minimization problem in (\ref%
{dualInitial}) is also%
\begin{equation*}
\forall x,y,~\tilde{u}^{\top }\left( x,y\right) +\tilde{v}^{\bot }\left(
x,y\right) \geq \Phi \left( x,y\right)
\end{equation*}%
which follows directly from the fact that%
\begin{equation*}
\forall x,\varepsilon ,y,\eta ,~\tilde{u}\left( x,\varepsilon \right) -\chi
\left( \left( x,\varepsilon \right) ,y\right) +\tilde{v}\left( y,\eta
\right) -\xi \left( x,\left( y,\eta \right) \right) \geq \Phi \left(
x,y\right) .
\end{equation*}
\end{enumerate}

Now define
\begin{equation*}
U\left( x,y\right) =\tilde{u}^{\bot }\left( x,y\right) \mbox{ and } V\left(
x,y\right)= \tilde{v}^{\bot }\left( x,y\right);
\end{equation*}
Given points 1. and 2. above, we can rewrite the value $\mathcal{W}$ as%
\begin{equation*}
\mathcal{W}=\inf_{(U,V) \in A }\left\{ \int U^{\top }\left( \tilde{x}\right)
d\tilde{P}\left( \tilde{x}\right) +\int V^{\top }\left( \tilde{y}\right) d%
\tilde{Q}\left( \tilde{y}\right)\right\}.
\end{equation*}
QED.
\end{varproof}

We are now in a position to prove the theorem.

\begin{varproof}[Proof of Theorem \protect\ref{theorem-socialw-primalp}]
Start by drawing two samples of size $N$ of men and women from their
population distributions $P$ and $Q$; we denote the corresponding values of
the observed characteristics $\left\{ x_{1},...,x_{n}\right\} $ and $\left\{
y_{1},...,y_{n}\right\} $. Call $P_{n}$ and $Q_{n}$ the corresponding sample
distributions; e.g. $P_{n}$ assigns a mass
\begin{equation*}
p_{i,n}=\frac{1}{n}\sum_{j=1}^{n}\mathrm{1\kern-.40em1}(x_{j}=x_{i}^{\ast })
\end{equation*}%
to the value $x_{i}^{\ast }$ of observable characteristics of men. The Law
of Large Numbers implies that $P_{n}$ and $Q_{n}$ converge in distribution
to $P$ and $Q$, the population distributions of the observable types. Now we
have for any possible $x$
\begin{equation*}
\int U^{\top }\left( \tilde{x}\right) d\tilde{P}_{n}\left( \varepsilon
|X=x\right) =\sum_{\substack{ i=1,..,n  \\ x_{i}=x}}\sup_{j=1,...,n}\left\{
U\left( x_{i},y_{j}\right) +\chi \left( \left( x_{i},\varepsilon _{i}\right)
,y_{j}\right) \right\} +o\left( 1\right)
\end{equation*}

As $N$ gets large enough, each of the possible values of observable
characteristics of women $y_{t}^{\ast }$ is included in the sample $\left\{
y_{1},...,y_{n}\right\} $; then the $\sup $ in the above expression runs
over all such possible values $\left\{ y_{1}^{\ast },...,y_{T_{y}}^{\ast
}\right\} $. But under (GUI), conditional on $X$ the random variables $\chi
\left( \left( x,\varepsilon \right) ,y_{t}^{\ast }\right) $ are independent
Gumbel random variables with scaling factor $\sigma _{1}$, so we get%
\begin{equation*}
\frac{1}{\sigma _{1}}\int U^{\top }\left( \tilde{x}\right) d\tilde{P}%
_{n}\left( \varepsilon |X=x\right) =\log \sum_{t=1}^{T_{y}}\exp \left(
U\left( x,y_{t}\right) /\sigma _{1}\right) +o_{P}\left( 1\right)
\end{equation*}%
hence, taking the limit and integrating over $x$,
\begin{equation*}
\int U^{\top }\left( \tilde{x}\right) d\tilde{P}\left( \tilde{x}\right)
=\sigma _{1}E_{P}\log \sum_{y}\exp \left( U\left( X,y\right) /\sigma
_{1}\right)
\end{equation*}%
and similarly
\begin{equation*}
\int V^{\top }\left( \tilde{y}\right) d\tilde{Q}\left( \tilde{y}\right)
=\sigma _{2}E_{Q}=\log \sum_{x}\exp \left( V\left( x,Y\right) /\sigma
_{2}\right) .
\end{equation*}%
QED.
\end{varproof}

\subsection{Proof of Theorem \protect\ref{theorem-socialw-primal}}

\begin{varproof}
By theorem (\ref{theorem-socialw-primalp}), we have
\begin{equation*}
\mathcal{W}_{N}=\inf_{U\left( x,y\right) +V\left( x,y\right) \geq \Phi
\left( x,y\right) \;\forall x,y}\left\{
\begin{array}{c}
\sigma _{1}\sum_{x}p\left( x\right) \log \left( \sum_{y}\exp \left( U\left(
x,y\right) /\sigma _{1}\right) \right) \\
+\sigma _{2}\sum_{y}q\left( y\right) \log \left( \sum_{x}\exp \left( V\left(
x,y\right) /\sigma _{2}\right) \right)%
\end{array}%
\right\}
\end{equation*}%
for which we form the Lagrangian%
\begin{eqnarray*}
\mathcal{W}_{N} &=&\inf_{U\left( x,y\right) ,V\left( x,y\right) }\sup_{\pi
\left( x,y\right) \geq 0}\left\{
\begin{array}{c}
\sigma _{1}\sum_{x}p\left( x\right) \log \left( \sum_{y}\exp \left( U\left(
x,y\right) /\sigma _{1}\right) \right) \\
+\sigma _{2}\sum_{y}q\left( y\right) \log \left( \sum_{x}\exp \left( V\left(
x,y\right) /\sigma _{2}\right) \right) \\
+\sum_{x,y}\pi \left( x,y\right) \left( \Phi \left( x,y\right) -U\left(
x,y\right) -V\left( x,y\right) \right)%
\end{array}%
\right\} \\
&=&\sup_{\pi \left( x,y\right) \geq 0}\left\{ \sum_{xy}\pi \left( x,y\right)
\Phi \left( x,y\right) +\inf_{U\left( .,.\right) }F(U)+\inf_{V\left(
.,.\right) }G(V)\right\}
\end{eqnarray*}%
where
\begin{eqnarray*}
F(U) &=&\sigma _{1}\sum_{x}p\left( x\right) \log \left( \sum_{y}\exp \left(
U\left( x,y\right) /\sigma _{1}\right) \right) -\sum_{xy}\pi \left(
x,y\right) U\left( x,y\right) \\
G(V) &=&\sigma _{2}\sum_{y}q\left( y\right) \log \left( \sum_{x}\exp \left(
V\left( x,y\right) /\sigma _{2}\right) \right) -\sum_{xy}\pi \left(
x,y\right) V\left( x,y\right) .
\end{eqnarray*}%
Clearly, $U\left( .,.\right) $ and $V\left( .,.\right) $ in the inner
minimization problems satisfy
\begin{equation}
\pi \left( x,y\right) =\frac{p\left( x\right) \exp \left( U\left( x,y\right)
/\sigma _{1}\right) }{\sum_{y}\exp \left( U\left( x,y\right) /\sigma
_{1}\right) }=\frac{q\left( y\right) \exp \left( V\left( x,y\right) /\sigma
_{2}\right) }{\sum_{x}\exp \left( V\left( x,y\right) /\sigma _{2}\right) };
\label{logitExpress}
\end{equation}%
note that these equations imply that $\sum_{y}\pi \left( x,y\right) =p\left(
x\right) $ and $\sum_{x}\pi \left( x,y\right) =q\left( y\right) $, so that $%
\pi \in \mathcal{M}(P,Q)$. Rearranging terms,
\begin{equation*}
\mathcal{W}_{N}=\sup_{\pi \in \mathcal{M}(P,Q)}\left\{
\begin{array}{c}
\sum_{xy}\pi \left( x,y\right) \Phi \left( x,y\right) -\left( \sigma
_{1}+\sigma _{2}\right) \sum_{xy}\pi \left( x,y\right) \log \pi \left(
x,y\right) \\
+\sigma _{1}\sum_{x}p\left( x\right) \log p\left( x\right) +\sigma
_{2}\sum_{y}q\left( y\right) \log q\left( y\right)%
\end{array}%
\right\}
\end{equation*}%
and noticing that $\sum_{xy}\pi \left( x,y\right) \log \pi \left( x,y\right)
=D\left( \pi \right) -S\left( P\right) -S\left( Q\right) $ gives the desired
result.
\end{varproof}

\subsection{Proof of Theorem \protect\ref{thm-socialw-homo}}

\begin{varproof}
The result follows directly from the Kantorovich duality (cf. %
\citep{Villani:2009}, Ch. 2); it can also be obtained by letting $%
\sigma _{1}$ and $\sigma _{2}$ tend to zero in Theorem \ref%
{theorem-socialw-primalp}, and noting that, as $\sigma _{1},\sigma
_{2}\rightarrow 0$,
\begin{eqnarray*}
\sigma _{1}E_{P}\left[ \log \sum_{y}\left[ \exp \left( U\left( X,y\right)
/\sigma _{1}\right) \right] \right] &\rightarrow &E_{P}\left[
\max_{y}U\left( X,y\right) \right] , \\
\sigma _{2}E_{Q}\left[ \log \sum_{x}\left[ \exp \left( V\left( x,Y\right)
/\sigma _{2}\right) \right] \right] &\rightarrow &E_{Q}\left[
\max_{x}U\left( x,Y\right) \right] .
\end{eqnarray*}
\end{varproof}

\subsection{Proof of Theorem \protect\ref{thm-crossders-ident}}

\begin{varproof}
(i) For $\sigma >0$, the map $\pi \rightarrow \sum_{x,y}\pi (x,y)\Phi \left(
x,y\right) -\sigma I\left( \pi \right) $ is strictly convave and finite, on
the convex domain $\mathcal{M}\left( P,Q\right) $; thus there exists a
unique $\pi \in \mathcal{M}\left( P,Q\right) $ maximizing (\ref{DualPT}).

(ii) Let $B$ be the set of pairs of functions $(u\left( x\right) ,v\left(
y\right) )$ such that $\sum_{x}u\left( x\right) p\left( x\right)
=\sum_{y}v\left( y\right) q\left( y\right) =0$, and for $\left( u,v\right)
\in B$, let $Z$ be the partition
\begin{equation*}
Z\left( u,v\right) :=\sum_{x,y}p\left( x\right) q\left( y\right) \exp \left(
\frac{\Phi \left( x,y\right) -u\left( x\right) -v\left( y\right) }{\sigma }%
\right) .
\end{equation*}%
Introduce
\begin{eqnarray*}
p_{u,v}\left( x\right) &:&=\frac{\partial \log Z\left( u,v\right) }{\partial
u\left( x\right) }=\frac{\sum_{y}p\left( x\right) q\left( y\right) \exp
\left( \frac{\Phi \left( x,y\right) -u\left( x\right) -v\left( y\right) }{%
\sigma }\right) }{\sum_{x,y}p\left( x\right) q\left( y\right) \exp \left(
\frac{\Phi \left( x,y\right) -u\left( x\right) -v\left( y\right) }{\sigma }%
\right) } \\
q_{u,v}\left( y\right) &:&=\frac{\partial \log Z\left( u,v\right) }{\partial
v\left( y\right) }=\frac{\sum_{x}p\left( x\right) q\left( y\right) \exp
\left( \frac{\Phi \left( x,y\right) -u\left( x\right) -v\left( y\right) }{%
\sigma }\right) }{\sum_{x,y}p\left( x\right) q\left( y\right) \exp \left(
\frac{\Phi \left( x,y\right) -u\left( x\right) -v\left( y\right) }{\sigma }%
\right) }
\end{eqnarray*}%
as a result $p_{u,v}$ and $q_{u,v}$ are probability vectors. By the strict
concavity of $\log Z$, there exists a unique vector $\left( u,v\right) \in B$
such that
\begin{eqnarray*}
p &=&p_{u,v} \\
q &=&q_{u,v}
\end{eqnarray*}%
and $\pi \left( x,y\right) =p\left( x\right) q\left( y\right) \exp \left(
\frac{\Phi (x,y)-u(x)-v(y)}{\sigma }\right) \in \mathcal{M}\left( P,Q\right)
$.

(iii) Let $\pi \in \mathcal{M}\left( P,Q\right) $ be the solution of (\ref%
{DualPT}). From Expression (\ref{logitExpress}) in the proof of Theorem \ref%
{theorem-socialw-primal}, we have that
\begin{eqnarray*}
\sigma _{1}\log \pi \left( x,y\right) &=&U\left( x,y\right) +\sigma _{1}\log
p\left( x\right) -\sigma _{1}\log \left( \sum_{y}\exp \left( U\left(
x,y\right) /\sigma _{1}\right) \right) \\
\sigma _{2}\log \pi \left( x,y\right) &=&V\left( x,y\right) +\sigma _{2}\log
q\left( y\right) -\sigma _{2}\log \left( \sum_{x}\exp \left( V\left(
x,y\right) /\sigma _{2}\right) \right)
\end{eqnarray*}%
thus, summing up%
\begin{equation*}
\sigma \log \frac{\pi \left( x,y\right) }{p\left( x\right) q\left( y\right) }%
=\Phi \left( x,y\right) -u\left( x\right) -v\left( y\right) -c
\end{equation*}%
where
\begin{eqnarray*}
u\left( x\right) &=&\sigma _{2}\log p\left( x\right) +\sigma _{1}\log \left(
\sum_{y}\exp \left( U\left( x,y\right) /\sigma _{1}\right) \right) +c_{1} \\
v\left( y\right) &=&\sigma _{1}\log q\left( y\right) +\sigma _{2}\log \left(
\sum_{x}\exp \left( V\left( x,y\right) /\sigma _{2}\right) \right) +c_{2} \\
c &=&c_{1}+c_{2}
\end{eqnarray*}%
and $c_{1}$ and $c_{2}$ are constant adjusted so that $\left( u,v\right) \in
B$. Hence $\pi $ is solution of equation (\ref{schrodinger-eq}). It follows
immediately that $c=\mathcal{W}$.
\end{varproof}

\subsection{Proof of Proposition \protect\ref{prop-cases}}

\begin{varproof}
a) The convexity of $\mathcal{W}$ follows from the fact that it is the
supremum of expression which are linear with respect to $\theta $.

b) As a result, by the enveloppe theorem, the subdifferential of $\mathcal{W}
$ at $\theta $ is the set of $\left\{ C(\Pi ),-I\left( \Pi \right) \right\} $
such that $\Lambda C(\Pi )-\sigma I\left( \Pi \right) =\mathcal{W}\left(
\Lambda ,\sigma \right) $. When this set consists of a single point, $%
\mathcal{W}$ is differentiable at $\theta $ and%
\begin{equation*}
\frac{\partial \mathcal{W}}{\partial \Lambda _{k}}(\theta )=C^{k}(\Pi ),~~~%
\frac{\partial \mathcal{W}}{\partial \sigma }(\theta )=-I\left( \Pi \right) .
\end{equation*}
\end{varproof}

\subsection{Proof of Proposition \protect\ref{prop-convexFeasible}}

\begin{varproof}
Non-emptiness is obvious. Now $\mathcal{F}_{c}$ is convex: Let $\hat{C}$ and
$\tilde{C}$ be two feasible cross-product matrices in $\mathcal{F}_{c}$. We
first show that for any $\alpha \in \left[ 0,1\right] $, $\alpha \hat{C}%
+\left( 1-\alpha \right) \tilde{C}$ is in $\mathcal{F}_{c}$. By definition
of $\mathcal{F}_{c}$, there exist $\hat{\pi}$ and $\tilde{\pi}$ in $\mathcal{%
M}\left( P,Q\right) $ such that $\hat{C}_{ij}=E_{\hat{\pi}}\left[
X_{ij}Y_{ij}\right] $ and $\tilde{C}_{ij}=E_{\tilde{\pi}}\left[ X_{ij}Y_{ij}%
\right] $. Let $\bar{\pi}=\alpha \hat{\pi}+\left( 1-\alpha \right) \tilde{\pi%
}$. Then $\alpha \hat{C}_{ij}+\left( 1-\alpha \right) \tilde{C}_{ij}=E_{\bar{%
\pi}}\left[ X_{ij}Y_{ij}\right] $, and $\bar{\pi}\in \mathcal{M}\left(
P,Q\right) $, thus $\alpha \hat{C}+\left( 1-\alpha \right) \tilde{C}\in
\mathcal{F}_{c}$. Now we prove that $\mathcal{F}_{c}$ is closed:\ Let $C_{n}$
be a sequence in $\mathcal{F}_{c}$ converging to $C\in \mathrm{I\kern-.17emR}%
^{rs}$, and let $\pi _{n}$ be the associated matching. By Theorem 11.5.4 in %
\citep{Dudley:02}, as $\mathcal{M}\left( P,Q\right) $ is uniformly
tight, $\pi _{n}$ has a weakly converging subsequence in $\mathcal{M}\left(
P,Q\right) $; call $\pi $ its limit. Then $C$ is the cross-product
associated to $\pi $, so that $C\in \mathcal{F}_{c}$. Finally, $\mathcal{F}$
is a closed convex set as it is the upper graph of the function $I_{r}\left(
C\right) $ defined in Eq. (\ref{IrC}).
\end{varproof}

\subsection{Proof of Proposition \protect\ref{prop-frontierFeasible}}

\begin{varproof}
$\mathcal{R}$ is the reunion of the subgradients of $\mathcal{W}$ which was
seen in Prop. \ref{prop-convexFeasible} to be the support function of $%
\mathcal{F}$: hence $\mathcal{R}$ is the frontier of $\mathcal{F}$.
\end{varproof}

\subsection{Proof of Proposition \protect\ref{prop-cases}}

\begin{varproof}
a) Positive homogeneity and convexity of degree one follows from the fact
that $\mathcal{W}$ is the support function of $\mathcal{F}$. Strict
convexity\ for $\sigma >0$ follows from the strict convexity of $I\left( \pi
\right) $. Part b) follows directly from the enveloppe theorem. Part c)
results of $I_{r}\left( C\right) $ being the Legendre transform of $\mathcal{%
W}\left( \lambda ,1\right) $ which is strictly convex, hence it convex on $%
\mathcal{F}_{c}$, differentiable on its interior, and by the enveloppe
theorem, $\frac{\partial I_{r}}{\partial C^{k}}=\frac{\Lambda _{k}}{\sigma }$%
.
\end{varproof}

\subsection{Proof of Proposition \protect\ref{prop-geometric}}

\begin{varproof}
a) The sets $\mathcal{R}_{c}\left( I\right) $\ are extreme points of the
sets $I_{r}^{-1}\left( \left[ 0,I\right] \right) $ which are closed convex
sets. One has $I_{r}^{-1}\left( \left\{ 0\right\} \right) =\left\{ C_{\infty
}\right\} $ which corresponds to $\Pi =P\otimes Q$, and $I_{r}^{-1}\left( %
\left[ 0,S\left( P\right) +S\left( Q\right) \right] \right) =\mathcal{F}_{c}$%
. Clearly, one has $\hat{C}\in \mathcal{R}_{c}\left( I_{r}\left( \hat{C}%
\right) \right) $. Finally, the form $\sum_{k}\Lambda _{k}dC^{k}$ vanishes
along $\mathcal{R}_{c}\left( I\right) $, so one has $\sum_{k}\Lambda
_{k}dC^{k}=0$, hence the result.
\end{varproof}

\section{Proof of Proposition \protect\ref{prop-partialsT}}

\begin{varproof}
By equation (\ref{schrodinger-eq}), we have%
\begin{equation*}
\log \frac{\pi \left( x,y\right) }{p\left( x\right) q\left( y\right) }=\frac{%
\Phi (x,y)-u(x)-v(y)-c}{\sigma }
\end{equation*}%
hence $\sigma \frac{\partial \log \pi }{\partial \Lambda _{k}}(x,y)=\phi
^{k}\left( x,y\right) -\frac{\partial u\left( x\right) }{\partial \Lambda
_{k}}-\frac{\partial v\left( x\right) }{\partial \Lambda _{k}}-\frac{%
\partial c}{\partial \Lambda _{k}}$. But we have that
\begin{equation*}
\sum_{x}\frac{\partial \log \pi _{\Lambda }\left( x,y\right) }{\partial
\Lambda _{k}}\pi _{\Lambda }\left( x,y\right) =\sum_{x}\frac{\partial \pi
_{\Lambda \left( x,y\right) }}{\partial \Lambda _{k}}=\frac{\partial }{%
\partial \Lambda _{k}}\sum_{x}\pi _{\Lambda \left( x,y\right) }=\frac{%
\partial q\left( y\right) }{\partial \Lambda _{k}}=0,
\end{equation*}%
thus for all $x$ and $y,$%
\begin{equation*}
E\left[ \frac{\partial \log \pi _{\Lambda }\left( X,Y\right) }{\partial
\Lambda _{k}}|X=x\right] =0\text{ and }E\left[ \frac{\partial \log \pi
_{\Lambda }\left( X,Y\right) }{\partial \Lambda _{k}}|Y=y\right] =0
\end{equation*}%
hence $\frac{\partial \log \pi }{\partial \Lambda _{k}}(x,y)\in V^{\circ }$,
while $\frac{\partial u\left( x\right) }{\partial \Lambda _{k}}+\frac{%
\partial v\left( y\right) }{\partial \Lambda _{k}}\in V^{+}$, therefore
\begin{equation*}
\phi ^{k}\left( x,y\right) =\sigma \frac{\partial \log \pi }{\partial
\Lambda _{k}}(x,y)+\frac{\partial u\left( x\right) }{\partial \Lambda _{k}}+%
\frac{\partial v\left( y\right) }{\partial \Lambda _{k}}+E\left[ \phi
^{k}\left( X,Y\right) \right]
\end{equation*}%
is the orthogonal decomposition of $\phi ^{k}\left( x,y\right) $ on $%
V^{\circ }\oplus V^{+}\oplus \mathbb{R}$, hence $\frac{\partial u\left(
x\right) }{\partial \Lambda _{k}}+\frac{\partial v\left( y\right) }{\partial
\Lambda _{k}}=\mathcal{P}\phi ^{k}\left( x,y\right) $.
\end{varproof}

\section{Proof of Proposition \protect\ref{hessian}}
\begin{varproof}
We have%
\begin{equation*}
\frac{\partial \mathcal{W}\left( \Lambda ,\sigma \right) }{\partial \Lambda
_{l}}=E\left[ \phi ^{l}\left( X,Y\right) \right]
\end{equation*}%
hence%
\begin{equation*}
\frac{\partial ^{2}\mathcal{W}\left( \Lambda ,\sigma \right) }{\partial
\Lambda _{k}\partial \Lambda _{l}}=E\left[ \phi ^{l}\left( X,Y\right) \frac{%
\partial \log \pi }{\partial \Lambda _{k}}(X,Y)\right] =\sigma E\left[ \frac{%
\partial \log \pi }{\partial \Lambda _{k}}(X,Y)\frac{\partial \log \pi }{%
\partial \Lambda _{l}}(X,Y)\right] .
\end{equation*}%
Further, by the orthogonality of $V^{\circ }$ and $V^{+}$,
\begin{eqnarray*}
cov\left( \phi ^{k}\left( X,Y\right) ,\phi ^{l}\left( X,Y\right) \right)
&=&\sigma ^{2}cov\left( \frac{\partial \log \pi }{\partial \Lambda _{k}}%
(X,Y),\frac{\partial \log \pi }{\partial \Lambda _{l}}(X,Y)\right) \\
&&+cov\left( \mathcal{P}\phi ^{k}\left( X,Y\right) ,\mathcal{P}\phi
^{l}\left( X,Y\right) \right)
\end{eqnarray*}%
QED.
\end{varproof}

\section{Proof of Theorem \protect\ref{thm-asymptotic}}

\begin{varproof}
We have $\hat{\lambda}_{N}=\frac{\partial I_{r}}{\partial C}$, hence at
first order $\hat{\lambda}_{N}-\lambda =D^{2}I_{r}.\left( \hat{C}%
_{N}-C\right) +o_{P}\left( 1/\sqrt{N}\right) $. But as $I_{r}$ is the
Legendre transform of $\mathcal{W}\left( \cdot ,1\right) $, it results that $%
D^{2}I_{r}=\left( D^{2}\mathcal{W}\left( \cdot ,1\right) \right) ^{-1}=%
\mathcal{I}^{-1}$ by Proposition \ref{hessian}.
\end{varproof}

\section{Connections to Statistical physics}

\label{physicsstuff} There is in fact, a very close parallel between our
theory and Statistical physics and Thermodynamics.\ We refer to %
\citep{Parisi:1988} for more on Statistical physics, and to %
\citep{MM:2009} for connection with Information theory. To give hints
to the parallel, let us just mention that the social welfare $\mathcal{W}$
is the analog of a \emph{total energy}; the term $\sum \lambda _{k}C^{k}$ is
the analog of an \emph{internal energy}; $I\left( \pi \right) $ is the
analog of an \emph{entropy}; the parameter $\sigma $ is the analog of a
\emph{temperature}. A pure matching is the equivalent of a \emph{solid state}%
; the points of nondifferentiability of $\mathcal{W}$ are analog to \emph{%
critical points}.

Note that equation \ref{schrodinger-eq} is known in the mathematical physics
literature as the Schr\"{o}dinger-Bernstein equation, cf. \citep{RT:98} and
references therein. It was first studied by Erwin Schr\"{o}dinger as part of
his research program in time irreversibility in Statistical Physics.
Interestingly, it also bears some connections with the better-known
\textquotedblleft Schr\"{o}dinger equation\textquotedblright\ in Quantum
mechanics of the same inventor. In fact, as discovered by Zambrini, a
dynamic formulation of this equation is the Euclidian Schr\"{o}dinger
equation which arises in Ed Nelson's formulation of \textquotedblleft
Stochastic Mechanics,\textquotedblright\ an Euclidian analog of quantum
mechanics. For more on this topic, see \citep{Parisi:1988}, Chap. 19.

\newpage


\markboth{References}{References}
\printbibliography

\end{document}